\documentclass[10pt,twocolumn,letterpaper]{article}

\usepackage{iccv}              

\usepackage[accsupp]{axessibility}  
\usepackage{float}
\usepackage{footnote}
\usepackage{enumitem}
\usepackage{bm}
\usepackage{arydshln}
\usepackage{booktabs}
\usepackage{multicol}
\usepackage{multirow}
\usepackage{color}
\usepackage{soul}
\usepackage{colortbl}
\usepackage{bbding}
\usepackage{makecell}
\usepackage{mathtools}
\usepackage{imakeidx}
\usepackage{pifont}
\usepackage{graphicx}
\usepackage{setspace}
\makeindex
\usepackage{arydshln}
\usepackage{lipsum}
\usepackage{natbib}
\usepackage[toc]{multitoc}
\usepackage[edges]{forest}
\usepackage[normalem]{ulem}
\definecolor{mydarkblue}{rgb}{0,0.08,0.45}

\usepackage{color}
\usepackage{colortbl}
\definecolor{flagship}{rgb}{0.93, 0.06, 0.41}

\usepackage{amsmath}

\usepackage{bbding}
\usepackage[most]{tcolorbox}

\usepackage{algorithm}
\usepackage{algorithmic}

\usepackage{minitoc}
\usepackage[toc,page,header]{appendix}

\usepackage[tikz]{bclogo}
\usepackage[framemethod=tikz]{mdframed}
\definecolor{bgblue}{RGB}{245,243,253}
\definecolor{ttblue}{RGB}{91,194,224}
\definecolor{mydarkblue}{rgb}{0,0.08,0.45}
\definecolor{darkgreen}{rgb}{0.0, 0.8, 0.0}

\newcommand{\method}{{\fontfamily{ppl}\selectfont
RadGPT}}
\newcommand{\dataset}{{\fontfamily{ppl}\selectfont
AbdomenAtlas 3.0}}

\newcommand{\numofdataset}{17}
\newcommand{\numofradiologist}{12}
\newcommand{\numofct}{9,262}
\newcommand{\numofslice}{2,789,975}
\newcommand{\numoftoken}{1,843,262}
\newcommand{\numofhospitals}{138}
\newcommand{\numofcountries}{19}

\newcommand{\numoflesionreport}{3,955}
\newcommand{\numofliverlesionreport}{1,472}
\newcommand{\numofpancreaticlesionreport}{361}
\newcommand{\numofkidneylesionreport}{2,122}

\newcommand{\numofliverlesioninstance}{5,582}
\newcommand{\numofpancreaticlesioninstance}{368}
\newcommand{\numofkidneylesioninstance}{4,424}

\newcolumntype{P}[1]{>{\centering\arraybackslash}p{#1}}
\newlength\savewidth

\definecolor{iccvblue}{rgb}{0.21,0.49,0.74}
\usepackage[pagebackref,breaklinks,colorlinks,allcolors=iccvblue]{hyperref}

\usepackage{threeparttable}

\def\paperID{12392} 
\def\confName{ICCV}
\def\confYear{2025}

\title{\method: Constructing 3D Image-Text Tumor Datasets}

\author{
Pedro R. A. S. Bassi\textsuperscript{1,2,3} \quad 
Mehmet Can Yavuz\textsuperscript{4} \quad
Ibrahim Ethem Hamamci \textsuperscript{6,7} \quad
Sezgin Er \textsuperscript{5} \\
Xiaoxi Chen\textsuperscript{8} \quad 
Wenxuan Li\textsuperscript{1} \quad
Bjoern Menze\textsuperscript{6} \quad
Sergio Decherchi\textsuperscript{3} \\
Andrea Cavalli\textsuperscript{2,3,9} \quad
Kang Wang\textsuperscript{4} \quad 
Yang Yang\textsuperscript{4} \quad 
Alan Yuille\textsuperscript{1} \quad 
Zongwei Zhou\textsuperscript{1,}\thanks{Correspondence to Zongwei Zhou (\href{mailto:zzhou82@jh.edu}{\textsc{zzhou82@jh.edu}})} \\[2.5mm]
\textsuperscript{1}Johns Hopkins University \quad
\textsuperscript{2}University of Bologna \quad
\textsuperscript{3}Italian Institute of Technology \\
\textsuperscript{4}University of California, San Francisco \quad
\textsuperscript{5}Istanbul Medipol University\quad
\textsuperscript{6}University of Zurich \\
\textsuperscript{7}ETH AI Center \quad
\textsuperscript{8}University of Illinois Urbana-Champaign \\
\textsuperscript{9}École Polytechnique Fédérale de Lausanne\\
[1.5mm]%
{\small Code, dataset, and models:~\href{https://github.com/MrGiovanni/RadGPT}{https://github.com/MrGiovanni/RadGPT}}
}

\begin{document}

\twocolumn[{%
\maketitle%
	\centering
	\includegraphics[width=1\textwidth, trim=0em 0em 0em 0em, clip]{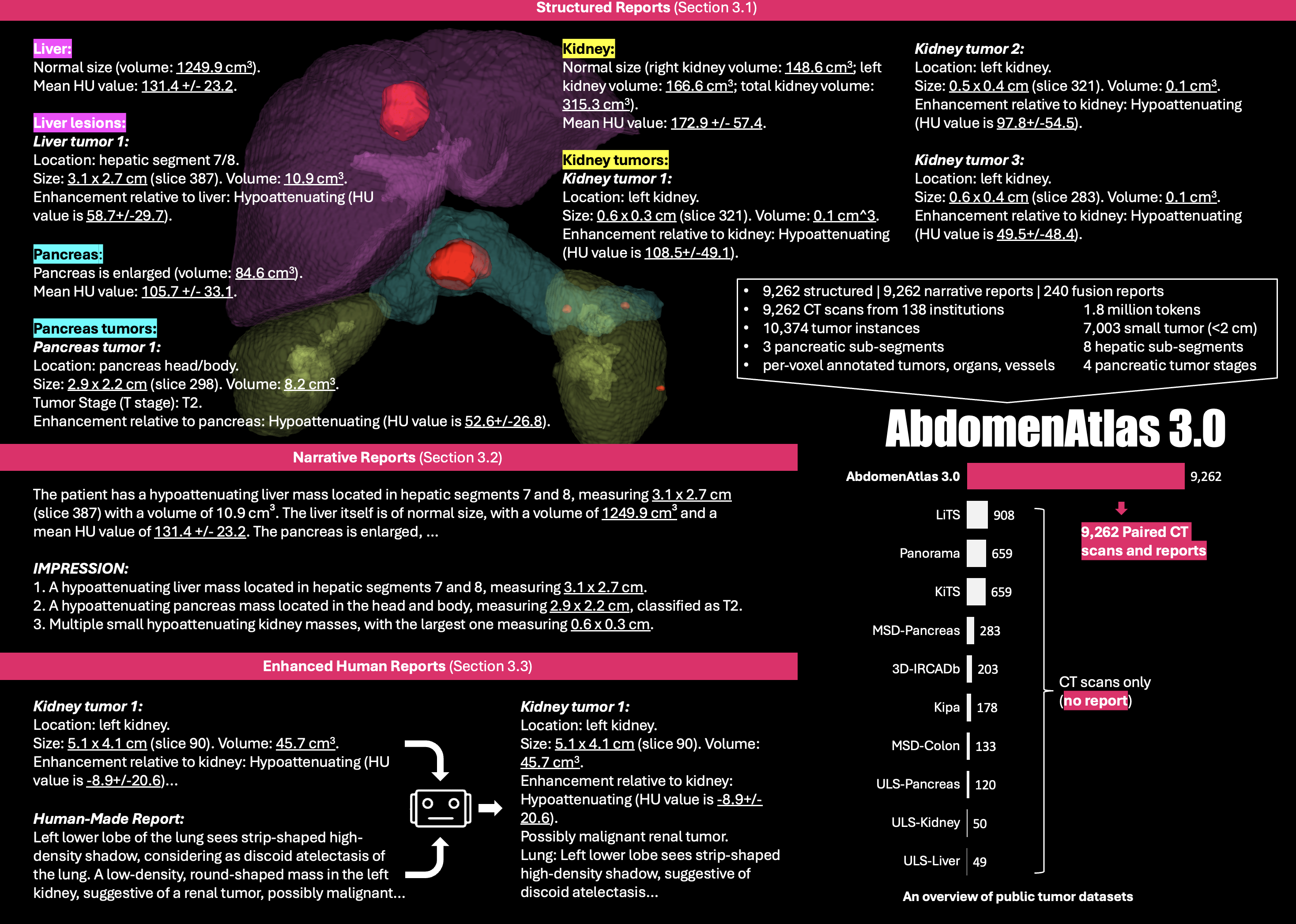}
    \captionof{figure}{\textbf{\dataset\ is a large-scale, image-text tumor dataset of 9,262 3D CT scans.} Each CT scan has per-voxel tumor annotations and reports, including \numofliverlesioninstance\ liver tumors, \numofpancreaticlesioninstance\ pancreatic tumors and \numofkidneylesioninstance\ kidney tumor, 7,003 of which are small tumors ($\leq$2cm). In addition, \dataset\ provides detailed annotations for pancreatic cancer staging (T1–T4), as well as per-voxel segmentation of liver sub-segments (1–8) and pancreatic sub-segments (head, body, and tail). Structured, narrative, and enhanced reports were created by a team of \numofradiologist\ board-certified radiologists assisted by our proposed Radiology Generative Pretrained Transformer (\method).
    }
    \label{fig:teaser}
 }]

\doparttoc 
\faketableofcontents 
\clearpage

\begin{abstract}

Cancers identified in CT scans are usually accompanied by detailed radiology reports, but publicly available CT datasets often lack these essential reports. This absence limits their usefulness for developing accurate report generation AI. 
To address this gap, we present \textbf{\dataset}, the first public, high-quality abdominal CT dataset with detailed, expert-reviewed radiology reports. All reports are paired with per-voxel masks and they describe liver, kidney and pancreatic tumors. \dataset\ has \numofct\ triplets of CT, mask and report---\numoflesionreport\ with tumors.
These CT scans come from 17 public datasets. Besides creating the reports for these datasets, we expanded their number of tumor masks by 4.2×, identifying 3,011 new tumor cases. 
Notably, the reports in \dataset\ are more standardized, and generated faster than traditional human-made reports. They provide details like tumor size, location, attenuation and surgical resectability. These reports were created by \textbf{12} board-certified radiologists using our proposed \textbf{\method}, a \textit{novel} framework that converted radiologist-revised tumor segmentation masks into structured and narrative reports. Besides being a dataset creation tool, \method\ can also become a fully-automatic, segmentation-assisted report generation method. We benchmarked this method and 5 state-of-the-art report generation vision-language models. Our results show that segmentation strongly improves tumor detection in AI-made reports.

\end{abstract}

\section{Introduction}\label{sec:introduction}

Each year, over 85 million CT scans are performed in the United States \cite{winder2021we,schockel2020developments}, growing 6\% per year, and significantly outpacing the 0.7\% annual growth rate of the medical imaging workforce \cite{codella2024medimageinsight}. This disparity puts radiologists under significant time pressure, making it challenging to generate detailed, accurate radiology reports. AI may support report generation, but it requires data. 
To address this gap, we present \dataset\ (summarized in \figureautorefname~\ref{fig:teaser} and \tableautorefname~\ref{tab:dataset_overview}), the first high-quality abdominal CT dataset with reports. It has \numofct\ 3D CTs in NIfTI format (\numofslice\ CT slices) sourced from \numofdataset\ public datasets (\tableautorefname~\ref{tab:dataset_overview}), which originally had no radiology report. \numofradiologist\ board-certified radiologists, assisted by \method\ (introduced below), generated reports for all CTs---totaling \numoftoken\ tokens. 
For each CT, we document tumor size, location, attenuation (HU), and volume for each identified tumor. Reports also include T-stage for pancreatic cancer (PDAC), derived from tumor size and vessel involvement, critical for surgery. Each CT has both structured (template-based) and narrative (free-text) reports, and precise voxel-level annotations. Reports cover tumors in the liver, pancreas, and kidneys, including 3,011 tumors newly identified by the radiologists.
Our reports also describe organ abnormalities (e.g., fatty liver, enlarged spleen), patient demographics, and contrast phase. They locate tumors in liver segments (1–8) and pancreas segments (head, body, tail)---all annotated per-voxel. This is the \ul{largest} liver sub-segment dataset, and the \ul{first} public pancreas sub-segment dataset.  Also, we enhanced 240 existing human-made reports, covering 66 distinct diagnoses, with more detailed tumor analyses.

\begin{table*}[t]
    \centering
    \scriptsize
    \begin{threeparttable}
    \begin{tabular*}{\textwidth}{@{\extracolsep{\fill}}p{0.2\linewidth}p{0.1\linewidth}p{0.1\linewidth}p{0.1\linewidth}p{0.1\linewidth}p{0.1\linewidth}p{0.1\linewidth}}
    \toprule
    dataset & CTs & institutions & countries & annotated\newline liver tumors & annotated\newline pancreatic tumors & annotated\newline kidney tumors \\
    \midrule
    FLARE'23 \citeyearpar{ma2022fast} [\href{https://codalab.lisn.upsaclay.fr/competitions/12239}{link}] & 4,100 & 35 & 1 & 0 $\rightarrow$ 564 & 0 $\rightarrow$ 38 & 0 $\rightarrow$ 941 \\
    KiTS'23 \citeyearpar{heller2020international} [\href{https://kits-challenge.org/kits23/}{link}] & 489 & 1 & 1 & 0 $\rightarrow$ 1 & 0 & 452 \\
    LiTS \citeyearpar{bilic2019liver} [\href{https://competitions.codalab.org/competitions/17094}{link}] & 131 & 7 & 5 & 50 & 0 & 0 \\
    TCIA-Pancreas-CT \citeyearpar{roth2015deeporgan} [\href{https://academictorrents.com/details/80ecfefcabede760cdbdf63e38986501f7becd49}{link}] & 42 & 1 & 1 & 0 & 0 & 0 \\
    CT-ORG \citeyearpar{rister2020ct} [\href{https://wiki.cancerimagingarchive.net/pages/viewpage.action?pageId=61080890#61080890cd4d3499fa294f489bf1ea261184fd24}{link}] & 140 & 8 & 6 & 0 $\rightarrow$ 44 & 0 & 0 $\rightarrow$ 21 \\
    Trauma Det. \citeyearpar{rsna-2023-abdominal-trauma-detection} [\href{https://www.rsna.org/education/ai-resources-and-training/ai-image-challenge/abdominal-trauma-detection-ai-challenge}{link}] & 4,714 & 23 & 13 & 0 $\rightarrow$ 113 & 0 $\rightarrow$ 32 & 0 $\rightarrow$ 38 \\
    BTCV \citeyearpar{landman2015miccai} [\href{https://www.synapse.org/#!Synapse:syn3193805/wiki/89480}{link}] & 47 & 1 & 1 & 0 & 0 & 0 \\
    CHAOS \citeyearpar{valindria2018multi} [\href{https://chaos.grand-challenge.org/Download/}{link}] & 20 & 1 & 1 & 0 & 0 $\rightarrow$ 1 & 0 \\
    AbdomenCT-1K \citeyearpar{ma2021abdomenct} [\href{https://github.com/JunMa11/AbdomenCT-1K}{link}] & 1,050 & 12 & 7 & 0 $\rightarrow$ 117 & 0 $\rightarrow$ 94 & 0 $\rightarrow$ 181 \\
    MSD CT Tasks (6) \citeyearpar{antonelli2021medical} [\href{https://decathlon-10.grand-challenge.org/}{link}] & 945 & 1 & 1 & 251 $\rightarrow$ 462 & 191 & 0 $\rightarrow$ 388 \\
    WORD \citeyearpar{luo2021word} [\href{https://github.com/HiLab-git/WORD}{link}] & 120 & 1 & 1 & 0 $\rightarrow$ 47 & 0 $\rightarrow$ 1 & 0 $\rightarrow$ 45 \\
    AMOS \citeyearpar{ji2022amos} [\href{https://amos22.grand-challenge.org}{link}] & 200 & 2 & 1 & 0 $\rightarrow$ 74 & 0 $\rightarrow$ 4 & 0 $\rightarrow$ 56 \\
    \textbf{\dataset\ (ours)} & \textbf{\numofct} & \textbf{\numofhospitals} & \textbf{\numofcountries} & \textbf{301 $\rightarrow$ \numofliverlesionreport} & \textbf{191 $\rightarrow$ \numofpancreaticlesionreport} & \textbf{452 $\rightarrow$ \numofkidneylesionreport} \\
    \midrule
    dataset & liver\newline sub-segments & pancreas\newline sub-segments & peripancreatic\newline blood vessels\footref{note:segmentations} & tumor\newline stage & radiology\newline reports & text\newline tokens \\
    \midrule
    FLARE'23 \citeyearpar{ma2022fast} [\href{https://codalab.lisn.upsaclay.fr/competitions/12239}{link}] & \ding{55} & \ding{55} & \ding{55} & \ding{55} & 0 & 0 \\
    KiTS'23 \citeyearpar{heller2020international} [\href{https://kits-challenge.org/kits23/}{link}] & \ding{55} & \ding{55} & \ding{55} & \ding{55} & 0 & 0 \\
    LiTS \citeyearpar{bilic2019liver} [\href{https://competitions.codalab.org/competitions/17094}{link}] & \textbf{\textcolor{darkgreen}{\ding{52}}} & \ding{55} & \ding{55} & \ding{55} & 0 & 0 \\
    TCIA-Pancreas-CT \citeyearpar{roth2015deeporgan} [\href{https://academictorrents.com/details/80ecfefcabede760cdbdf63e38986501f7becd49}{link}] & \ding{55} & \ding{55} & \ding{55} & \ding{55} & 0 & 0 \\
    CT-ORG \citeyearpar{rister2020ct} [\href{https://wiki.cancerimagingarchive.net/pages/viewpage.action?pageId=61080890#61080890cd4d3499fa294f489bf1ea261184fd24}{link}] & \ding{55} & \ding{55} & \ding{55} & \ding{55} & 0 & 0 \\
    Trauma Det. \citeyearpar{rsna-2023-abdominal-trauma-detection} [\href{https://www.rsna.org/education/ai-resources-and-training/ai-image-challenge/abdominal-trauma-detection-ai-challenge}{link}] & \ding{55} & \ding{55} & \ding{55} & \ding{55} & 0 & 0 \\
    BTCV \citeyearpar{landman2015miccai} [\href{https://www.synapse.org/#!Synapse:syn3193805/wiki/89480}{link}] & \ding{55} & \ding{55} & \ding{55} & \ding{55} & 0 & 0 \\
    CHAOS \citeyearpar{valindria2018multi} [\href{https://chaos.grand-challenge.org/Download/}{link}] & \ding{55} & \ding{55} & \ding{55} & \ding{55} & 0 & 0 \\
    AbdomenCT-1K \citeyearpar{ma2021abdomenct} [\href{https://github.com/JunMa11/AbdomenCT-1K}{link}] & \ding{55} & \ding{55} & \ding{55} & \ding{55} & 0 & 0 \\
    MSD CT Tasks (6) \citeyearpar{antonelli2021medical} [\href{https://decathlon-10.grand-challenge.org/}{link}] & \ding{55} & \ding{55} & \ding{55} & \ding{55} & 0 & 0 \\
    WORD \citeyearpar{luo2021word} [\href{https://github.com/HiLab-git/WORD}{link}] & \ding{55} & \ding{55} & \ding{55} & \ding{55} & 0 & 0 \\
    AMOS \citeyearpar{ji2022amos} [\href{https://amos22.grand-challenge.org}{link}] & \ding{55} & \ding{55} & \ding{55} & \ding{55} & 0 & 0 \\
    \textbf{\dataset\ (ours)} & \textbf{\textcolor{darkgreen}{\ding{52}}} & \textbf{\textcolor{darkgreen}{\ding{52}}} & \textbf{\textcolor{darkgreen}{\ding{52}}} & \textbf{\textcolor{darkgreen}{\ding{52}}} & \textbf{18,524} & \textbf{\numoftoken} \\
    \bottomrule
    \end{tabular*}
    \begin{tablenotes}
        \footnotesize
        \item $\rightarrow$ represents the number of CT scans with tumor annotations in the original dataset, followed ($\rightarrow$) by our updated number of CT scans with tumor annotations, including the additional annotations \dataset\ provided with radiologist support.
    \end{tablenotes}
    \end{threeparttable}
    \caption{\textbf{Besides being the only public abdominal CT dataset with paired radiology reports, \dataset\ offers 4.2$\times$ more annotated tumors than the combined total of its constituent datasets}. The table highlights how \dataset\ enhances public datasets with reports and tumor annotations. It includes \numofliverlesionreport\ CT scans with liver tumors, \numofpancreaticlesionreport\ with pancreatic tumors, and \numofkidneylesionreport\ with kidney tumors, most newly annotated with radiologist support. Each sample includes per-voxel annotations and reports. \dataset\ is also the first dataset to provide per-voxel segmentations of pancreas sub-segments and peripancreatic blood vessels. AbdomenAtlas 1.1 \cite{li2024abdomenatlas} has the same CTs as AbdomenAtlas 3.0, but it has only organ segmentation masks---no tumor masks, reports, organ sub-segments, nor peripancreatic blood vessels. AbdomenAtlas 2.0 has the same CTs and masks as 3.0, no report---it is our intermediate step before 3.0.
    }
    \label{tab:dataset_overview}
\end{table*}

To create \dataset, we developed Radiology Generative Pre-trained Transformers (\textbf{\method}), an anatomy‑aware vision‑language AI agent that assists radiologists in creating CT-report datasets. We started with our previous \textbf{AbdomenAtlas 1.1} \cite{li2024abdomenatlas}, composed of 17 public datasets and their organ segmentation masks, but no tumor segmentation nor report. First, \method\ segments liver, kidney, and pancreas tumors, along with liver/pancreas sub-segments, surrounding organs, and blood vessels\textsuperscript{\ref{note:segmentations}}. Then, radiologists revise the segmented tumors, annotating missed ones and removing false positives. We call the dataset with CT scans and tumor segmentation masks \textbf{AbdomenAtlas 2.0}, and we also release it here. From the revised segmentations, \method\ extracts attributes (e.g., tumor size, volume, attenuation, stage) via deterministic, rule-based algorithms. These attributes are used to fill a radiologist-designed template, producing \textit{structured reports}. \method's deterministic algorithms ensure that the structured reports are fully explainable and fully coherent with the radiologist-revised segmentations. Next, \method\ converts the structured reports into free-text \textit{narrative reports}, using large language models (LLMs) that emulate the style (word choice and organization) of radiologists at a major US hospital---through in-context learning with special example selection (\S\ref{sec:style}). Last, \method\ fuses per-voxel segmentations with human-made reports/clinical notes to produce \emph{enhanced human reports} (\S\ref{sec:fusion_report}), combining precise and detailed tumor analysis from segmentation with broader diagnostic range (66 diagnoses) from human-made reports. Reports were verified by radiologists (Appendix \ref{ap:radiologist_verification}). We call the final triplet dataset---CT scans, tumor masks, reports---\textbf{AbdomenAtlas 3.0}.

We evaluated six CT report generation models on \dataset\ (internal validation) and a private dataset (external validation): CT2Rep \cite{hamamci2024ct2rep}, M3D \cite{bai2024m3d}, CT-CHAT \cite{hamamci2024foundation}, Merlin \cite{blankemeier2024merlin}, RadFM \cite{wu2023towards} and \method. Besides a dataset creation tool, \method\ can also become a fully-automatic, segmentation-assisted report generation model, by converting the outputs of a segmentation model into reports, without radiologist revision. We expect \dataset\ to foster segmentation-assisted report generation, as the dataset has CTs, per-voxel annotations and reports. We evaluated all report generation models with a new diagnostic metric (\S\ref{sec:results_metric}). It first uses an LLM to extract labels (tumor presence) from AI- and human-made reports. Then, it compares the labels from AI- and human-made reports to evaluate AI's diagnostic sensitivity and specificity (\S\ref{sec:metric}). To validate this new metric, radiologists manually evaluated LLM labeling---it achieved 96\% zero-shot accuracy (\figureautorefname~\ref{fig:confusion}). Our contributions are:

\begin{enumerate}
    \item AbdomenAtlas 3.0 is the first public dataset with high-quality abdominal CT scans (9,262), radiology reports (structured, narrative, and enhanced), and tumor masks.
    \item With 12 radiologists, we annotated 3,011 new tumors in the 17 public datasets inside AbdomenAtlas 3.0—expanding their number of tumor masks by 4.2×.
    \item Our reports locate liver and pancreas tumors within sub-segments of the organs. They also measure contact between tumors and blood vessels for pancreatic tumor staging. Staging and sub-segments are key for surgery.
    \item We developed Rad-GPT to assist dataset creation: unlike current VLMs, it uses deterministic algorithms to convert radiologist-revised tumor masks into reports, improving reports’ trustworthiness and interpretability. Also, RadGPT can generate fully-automated reports.
    \item We benchmarked 5 SOTA VLMs for report generation and showed segmentation improves report generation.
\end{enumerate}


\begin{figure*}[t]
    \centering
    \includegraphics[width=1\linewidth]{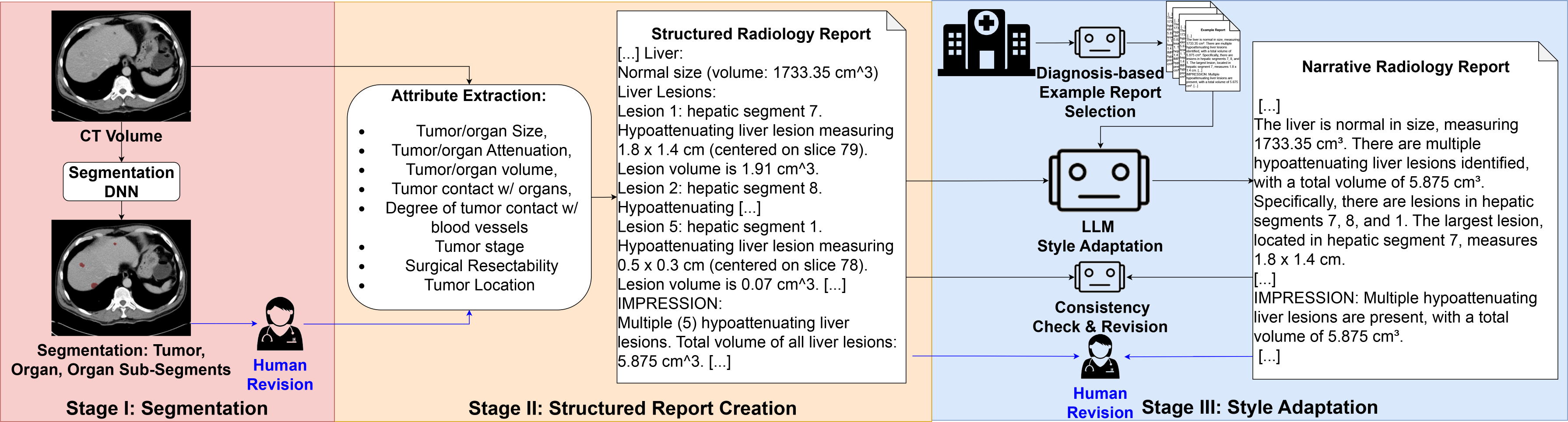}
    \caption{
    \textbf{The \method\ 3-stage pipeline for report generation.} Blue arrows denote human revision used to create \dataset.
    By skipping revision, \method\ can also become a fully-automatic, segmentation-assisted report generation AI.
    \textbf{Stage I. Segmentation.} DiffTumor \cite{chen2024towards} and nnU-Net \cite{isensee2021nnu} segment 26 anatomical structures important for cancer detection and staging\footref{note:segmentations}. Radiologists corrected wrong tumor segmentations in \dataset, and ground-truths from public datasets were used when available.
    \textbf{Stage II. Structured Report Generation.} Deterministic algorithms (\S\ref{sec:tumor-localization}-\ref{sec:tumor-staging}) extract radiologist-selected attributes---important for cancer detection, staging and treatment---from CTs and segmentations. Attributes fill a radiologist-designed template, generating structured reports detailing liver, kidney, and pancreatic tumors. The rule-based deterministic algorithms ensure the reports are fully coherent with segmentations and explainable.
    \textbf{Stage III. Style Adaptation.} LLM adapts structured reports into a target hospital’s narrative style, leveraging example reports from the hospital---in-context learning prioritizing examples of similar diagnoses (\S\ref{sec:style}). LLM is asked to preserve medical information and double checks for consistency. Radiologists revised reports in \dataset. Also, LLM can fuse structured and human-made reports, creating enhanced human reports combining segmentation-based precision with humans' broad diagnostic range (\S\ref{sec:fusion_report}).
    }
    \label{fig:rgpt}
\end{figure*}

\section{Related Work}
\label{sec:related_work}

Per-voxel tumor annotations are scarce. Most public abdominal CT datasets concentrate on a single tumor type (e.g., liver \cite{bilic2019liver}, pancreas \cite{antonelli2022medical}, or kidney \cite{heller2019kits19}) and contain only a few hundred tumor annotations (\tableautorefname~\ref{tab:dataset_overview}). This small volume of annotations hinders effective AI training and evaluation. To address this, our radiologists have \ul{quadrupled the number of per-voxel tumor annotations} in the \numofdataset\ public datasets included in \dataset\ (\tableautorefname~\ref{tab:dataset_overview}).

Real-world radiology reports are even rarer than per-voxel tumor annotations. At the time of writing, no publicly available abdominal CT dataset contains authentic clinical reports. Only one dataset, M3D-Cap \cite{bai2024m3d}, provides textual captions (sourced from Radiopaedia \cite{gaillard2011radiopaedia}), \ul{but its scans are 2D JPG/PNG image series rather than standard 3D NIfTI or DICOM volumes.} Consequently, crucial information such as inter-slice spacing and Hounsfield units (HU) is missing \cite{zhou2022interpreting}. In contrast, CT scans in \dataset\ were collected in standard formats from \numofhospitals\ medical institutions, retaining clinically important metadata. As another unique quality, our reports are paired with tumor and organ masks---fostering the development of segmentation-assisted report generation AI.

Due to the scarcity of reports in public datasets, only two models specifically target abdominal CT report generation: M3D \cite{bai2024m3d} (publicly released) and Merlin \cite{blankemeier2024merlin} (partially released). Text-similarity metrics were used to evaluate both models (e.g., BLEU and ROUGE \cite{lin2004rouge}; Merlin was also evaluated with RadGraph-F1), \ul{but these metrics can be skewed by style variations even when the underlying diagnoses remain unchanged} (\S\ref{sec:results_metric}). In contrast, we propose the evaluation of AI-generated reports using diagnostic sensitivity and specificity (\tableautorefname~\ref{tab:radGPT})---clinically meaningful and acceptable metrics \cite{xia2022felix,cao2023large}.
Lastly, although many report-generation models exist for 2D X-ray \cite{wang2018cvpr,chen2020miccai,li2019miccai,li2022cvpr,zhang2022eccv,tang2023iccv}, adapting them to 3D CT may require profound re-design, which may unfairly represent the originals. \textit{Why?} \textbf{\emph{First}}, tumors in CTs can occupy $\leq$0.0001\% of the full volume, vs. \ 5–10\% in X-rays. \textbf{\emph{Second}}, many X-ray models rely on 2D pre-trained models, but CT data is 3D. Processing CT slices individually is computationally prohibitive, and it is difficult to align slices with findings in reports. Thus, all models we evaluated in \dataset\ \cite{hamamci2024ct2rep,bai2024m3d,hamamci2024foundation,blankemeier2024merlin,wu2023towards} are designed for CT.

\section{\dataset\ \& \method}\label{sec:dataset}



\tableautorefname~\ref{tab:dataset_overview} shows advantages of \textbf{\dataset} over its \numofdataset\ source datasets---\textit{providing reports, organ sub-segments and blood vessels annotated per-voxel, and 4$\times$ more tumor annotations}. Sections \S\ref{sec:structured_report}--\S\ref{sec:fusion_report} explain \method\ (summarized in \figureautorefname~\ref{fig:rgpt}), and how it empowered 12 radiologists to generate reports for the \numofct\ CTs in \dataset.


\subsection{Creating Structured Reports}\label{sec:structured_report}

Structured reports use a radiologist-designed template, enhancing clarity and aiding medical decisions \cite{al2014pancreatic} (\figureautorefname~\ref{fig:teaser}). To fill the template, \method\ uses segmentation and deterministic algorithms to:
(1) sub-segment organs to locate tumors (\S\ref{sec:tumor-localization});
(2) measure tumor size, volume, and attenuation (\S\ref{sec:tumor-measurement});
(3) perform cancer staging from tumor and blood vessel segmentations (\S\ref{sec:tumor-staging}).

\subsubsection{Sub-segment Organs to Locate Tumors}
\label{sec:tumor-localization}

Human-made reports use organ sub-segments to locate tumors. Location is key for prognosis, tracking tumor progression, and treatment planning. E.g., the possibility of tumor surgical removal depends on its location \cite{tomasello2019outcome}. To locate liver and pancreas tumors in structured reports, \method\ sub-segments the organs and checks which sub-segments intersect with the tumor. \method\ segments tumors with DiffTumor \cite{chen2024towards}, a public segmentation model, and radiologists revise the segmentations (Appendix \ref{ap:radiologist_verification}).

For liver sub-segmentation, we leverage whole-liver ground-truth per-voxel annotations to help the AI find liver sub-segments. First, we offset the liver intensity (by 200 HU), following its ground-truth per-voxel annotation. Second, using these CT scans with offsets as input, we trained an nnU-Net \cite{isensee2021nnu} for liver sub-segmentation. The sub-segments follow the Couinaud standard \cite{couinaud1957foie}, which divides the liver into eight sub-segments that can be independently removed in surgeries. Couinaud sub-segment annotations are publicly available for 131 LiTS CT scans \cite{zhang2024robust,bilic2019liver}, which we used for training. Given the small size of this dataset, we fine-tuned an nnU-Net pre-trained on 9,262 CT scans in AbdomenAtlas 1.1 \cite{li2024abdomenatlas}. After fine-tuning, we inferenced the nnU-Net on \dataset. The HU value offsetting ensured the precise alignment between the generated sub-segments and the existing liver ground-truth per-voxel annotations. \dataset\ is the second \cite{zhang2024robust} but \textit{largest public dataset with liver sub-segments}.

For pancreas sub-segmentation, there is no public dataset with per-voxel annotations of pancreas head, body, and tail---ours is the first. Thus, to subsegment the pancreas, we used the superior mesenteric artery (SMA) as a landmark. We trained an nnU-Net to segment the SMA (using private data) and developed a deterministic algorithm that uses the SMA segmentation to sub-segment the pancreas (Sup. Alg. \ref{algo:pancreas_segmentation}). First, it uses the SMA to find the pancreatic neck, since it curves around the SMA. The neck locates the head-body boundary. Then, the body-tail boundary is set at the midpoint along their length. Our landmark-based deterministic algorithm closely mimics how radiologists use mesenteric vessels to subsegment the pancreas \cite{triay2022pancreas}. \dataset\ is the \textit{first public dataset with pancreas sub-segments}.

\subsubsection{Measure Tumors Like Radiologists}
\label{sec:tumor-measurement}

Radiologists commonly measure tumors using use the World Health Organization (WHO) standard, which provides two diameters: the longest tumor diameter in any axial plane ($D$), and its perpendicular diameter in the same plane ($d$) \cite{miller1981reporting}. Standardization of measurements is key for accurate cancer prognosis and treatment planning \cite{miller1981reporting,li2025scalemai}. Thus, \method\ also uses the WHO standard, measuring tumors like radiologists. \dataset\ presents radiologist-revised segmentations of liver, kidney and pancreas tumors. From segmentations, \method\ extracts tumor measurements using a deterministic algorithm that implements the WHO standard (Sup. Alg.~\ref{algo:measurement}). Besides diameters, our structured reports present tumor \& organ \textit{volume} and \textit{attenuation} (HU values), also extracted from segmentation\footnote{\dataset\ is the first dataset with per-voxel annotations for the blood vessels key for pancreatic tumor staging: the celiac axis (CA), superior mesenteric artery (SMA), superior mesenteric vein (SMV), common hepatic artery (CHA), and portal vein. These annotations were produced by an nnU-Net trained in private data, and revised by radiologists (Appendix \ref{ap:radiologist_verification}). \dataset\ also has per-voxel annotations for other 22 structures important for cancer detection/staging: liver tumors, kidney tumors, pancreas tumors, liver, kidney, pancreas, spleen, adrenal glands, stomach, duodenum, bile duct, intestines, aorta, and postcava. \label{note:segmentations}}. Using volumes, our reports diagnose enlarged organs, and attenuation diagnoses fatty liver (average HU $<40$ \cite{kodama2007comparison}) and pancreas (pancreas-to-spleen attenuation $< 0.7$ \cite{fukuda2017ct})---a condition related to diabetes and pancreatic cancer \cite{fukuda2017ct}. Meanwhile, tumor attenuation helps identify tumor type.

\begin{figure}[t]
    \centering
    \includegraphics[width=1\linewidth]{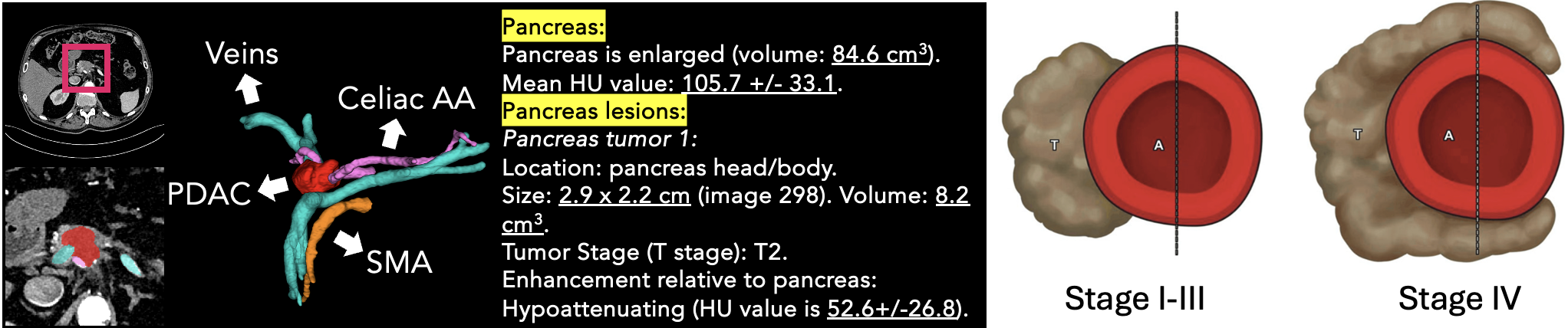}
    \caption{
    \textbf{Automated T staging.} Our \method\ first segments the tumor and key vascular structures from CT scans, then measures tumor size and blood vessel contact angle to automatically assign T stage and resectability. If the tumor-vessel contact angle surpasses 180 degrees, the tumor becomes unresectable (T stage 4).
    }
    \label{fig:antomated_T_staging}
\end{figure}

\subsubsection{Stage Pancreatic Cancer using Segmentation}
\label{sec:tumor-staging}

Tumor T‑stage summarizes tumor size and relationship to nearby structures. It is key for surgical planning and survival, especially for pancreatic adenocarcinoma (PDAC), an aggressive cancer \cite{al2014pancreatic}. However, staging is time-consuming. As shown in \figureautorefname~\ref{fig:antomated_T_staging}, for PDAC staging, radiologists must measure tumors (\S\ref{sec:tumor-measurement}) and analyze its interaction with blood vessels (SMA, CHA, CA, SA) \cite{al2014pancreatic}. Accordingly, \method\ first segments vessels and tumors (using nnU-Net and DiffTumor\footref{note:segmentations}) and radiologists revise segmentations (Appendix \ref{ap:radiologist_verification}). Then, a deterministic algorithm uses the revised segmentations to measure the tumor-vessel contact angle (Sup. Alg.~\ref{algo:staging}). Large angles ($>$180$^{\circ}$) make surgery difficult, increasing stage. For interpretability, reports justify stages with tumor size and tumor-vessel degree of contact, and our deterministic algorithm faithfully implements the guidelines radiologists use to stage PDAC \cite{al2014pancreatic}. \dataset\ is \textit{first public dataset with PDAC T stage labels}. 

\begin{table*}[t]
    \centering
    \scriptsize
    \begin{tabular}{p{0.09\linewidth}P{0.08\linewidth}P{0.08\linewidth}P{0.06\linewidth}P{0.08\linewidth}P{0.08\linewidth}P{0.06\linewidth}P{0.08\linewidth}P{0.08\linewidth}P{0.06\linewidth}}
    \multicolumn{10}{l}{\textit{Internal validation on the test set of \dataset\ (IID)}} \\
    \toprule
    & \multicolumn{3}{c}{pancreatic tumor (\%)} & \multicolumn{3}{c}{kidney tumor (\%)} & \multicolumn{3}{c}{liver tumor (\%)} \\
    \cmidrule(lr){2-4} \cmidrule(lr){5-7} \cmidrule(lr){8-10}
    Model & Sen.~($\leq$2~cm) & Sen.~($>$2~cm) & Spec. & Sen.~($\leq$2~cm) & Sen.~($>$2~cm) & Spec. & Sen.~($\leq$2~cm) & Sen.~($>$2~cm) & Spec. \\
    \midrule
CT-CHAT \cite{hamamci2024foundation} 
    & \textbf{66.7} & 51.9 & 61.2
    & 31.1 & 32.8 & 74.2
    & 5.7 & 3.2 & 94.7 \\
CT2Rep \cite{hamamci2024ct2rep} 
    & 0.0 & 0.0 & 92.5
    & 36.5 & 39.3 & 70.4
    & 35.8 & 49.2 & 70.4 \\
M3D \cite{bai2024m3d} 
    & 0.0 & 7.4 & 97.2
    & 8.1 & 16.4 & 84.1
    & 9.4 & 12.7 & 86.0 \\
Merlin \cite{blankemeier2024merlin} 
    & 33.3 & 51.9 & 71.8
    & 28.4 & 45.9 & 86.6
    & 30.2 & 41.3 & \textbf{95.9} \\
RadFM \cite{wu2023towards} 
    & 0.0 & 0.0 & \textbf{99.9}
    & 3.7 & 6.3 & \textbf{95.6}
    & 3.3 & 5.7 & 93.9 \\
\method\ (ours) 
    & \textbf{66.7} & \textbf{81.5} & 93.2
    & \textbf{54.8} & \textbf{93.3} & 51.8
    & \textbf{39.6} & \textbf{96.8} & 64.4 \\
\bottomrule
    \vspace{0.5em}\\
    \multicolumn{10}{l}{\textit{External validation on unseen hospital---UCSF (OOD)}} \\
    \toprule
    Model & Sen.~($\leq$2~cm) & Sen.~($>$2~cm) & Spec. & Sen.~($\leq$2~cm) & Sen.~($>$2~cm) & Spec. & Sen.~($\leq$2~cm) & Sen.~($>$2~cm) & Spec. \\
    \midrule
CT-CHAT \cite{hamamci2024foundation} 
    & 27.5 & N/A & 73.1
    & 24.3 & 29.7 & 74.6
    & 5.2 & 4.2 & 94.0 \\
CT2Rep \cite{hamamci2024ct2rep}
    & 2.1 & N/A & 96.7  
    & 4.0 & 10.0 & 98.0 
    & 0.0 & 0.0 & \textbf{100.0} \\  
M3D \cite{bai2024m3d} 
    & 3.3 & N/A & 97.9
    & 14.8 & 13.1 & 86.3
    & 10.7 & 17.3 & 87.3 \\
Merlin \cite{blankemeier2024merlin} 
    & 7.5 & N/A & \textbf{100.0}
    & 8.1 & 9.2 & \textbf{100.0}
    & 9.1 & 19.2 & \textbf{100.0} \\
RadFM \cite{wu2023towards} 
    & 0.0 & N/A & \textbf{100.0}
    & 7.5 & 6.8 & 90.9
    & 10.9 & 11.1 & 85.0 \\
\method\ (ours)
    & \textbf{76.9} & N/A & 76.6  
    & \textbf{92.0} & \textbf{97.3} & 78.3  
    & \textbf{79.6} & \textbf{89.4} & 73.4 \\  
\bottomrule
    \end{tabular}
    \caption{\textbf{In tumor detection, fully-automated reports by \method\ surpass reports created by end-to-end report generation models.} We use \method\ as a fully-automated segmentation-assisted report generation model (\figureautorefname~\ref{fig:rgpt}). The results indicate that per-voxel segmentation (step 1 in the \method\ pipeline) may strongly improve report generation. We tested out-of-distribution (OOD) at UCSF, a hospital not seen in training, and in-distribution (IID). In the IID set, the ground-truth contained 9 small and 27 large pancreatic tumors, 74 small and 61 large kidney tumors, and 53 small and 63 large liver tumors, with 890, 791, and 810 negatives, respectively. In OOD, we have 385 (small) and 0 (large) for pancreas, 50 (small) and 219 (large) for kidney, and 142 (small) and 301 (large) for liver, with 244 negatives for each organ. Decision thresholds are analyzed in \ref{fig:th}. While other methods were evaluated zero-shot, CT-CHAT, CT2Rep and Merlin were trained in \dataset, giving them an advantage in the IID dataset. To compute sensitivity and specificity, we used our proposed diagnostic evaluation (\S\ref{sec:metric}): an LLM extracted binary tumor presence labels \textit{per-organ}, and we compared the labels for AI-made reports and ground-truth human-made reports. LLM label extraction accuracy is 96\% (Figure \ref{fig:confusion}). Table \ref{tab:LLM_metrics} provides additional metrics (BLEU, ROUGE, BERT, RadGraph-F1), showing they are usually sensible to variations in report style, unlike our diagnostic evaluation.}
    \label{tab:radGPT}
\end{table*}



\subsection{Creating Narrative Reports}
\label{sec:style}

Structured reports use rigid templates to improve clarity and clinical decision-making \cite{al2014pancreatic}. However, rigid templates may conflict with the reporting style of an institution. Thus, \method\ can create narrative reports that mimic the style of a target institution. In \dataset, they mimic human-made reports at a major US hospital (\figureautorefname~\ref{fig:teaser}). The narrative reports are created through style adaptation with in-context learning: we provide a pre-trained LLM (Llama-3.1 70B, AWQ quantization \cite{dubey2024llama3herdmodels}) with a structured report and 10 human-made reports from the target institution, and the LLM adapts the structured report to the style of the human-made reports. We ask the LLM \textit{not} to change diagnoses or details. Thus, narrative reports contain all the detailed information from structured reports (\S\ref{sec:structured_report}).

However, style of human-made reports varies with diagnoses. E.g., pancreatic tumor differ from liver tumor reports \cite{RSNA2025RadReport}. Thus, we verify diagnoses to give the LLM example reports with the correct style. First, another LLM categorizes human-made reports according to tumors (liver, pancreas, kidney, none). Then, when adapting a structured report to narrative, the first LLM receives example human-made reports with the same tumor as the structured report. 

After adapting a structured report into a narrative report, the LLM performed a quality check. It extracted diagnoses and quantitative information (e.g., tumor size and stage) from both reports and checked for consistency. We prompted the LLM to correct in the narrative report any information diverging from the structured report, and to remove any diagnosis not present in the structured report.

\subsection{Creating Enhanced Human Reports}\label{sec:fusion_report}

Like most abdominal CT datasets, \dataset\ focuses on tumors---as cancer is a major cause of death. Our reports can precisely measure and analyze multiple tumors in a CT, while human-made reports usually measure the largest tumors only (\S \ref{sec:fusion_results}). However, human-made reports cover multiple diagnoses unrelated to tumors. To combine their strengths, \method\ prompts the zero-shot LLM (Llama 3.1 70B AWQ) to fuse the details in structured reports with the many diagnoses in human-made reports/clinical notes (Figures \ref{fig:fusion_TS} and \ref{fig:teaser}), generating \textit{enhanced human reports}. \dataset\ has 240 of them: 209 used clinical notes for TotalSegmentator CT scans \cite{bai2024m3d}, and 31 used notes from our radiologists. They span 66 diagnoses.

\subsection{Evaluating Diagnoses in AI-Made Reports}
\label{sec:metric}


We propose a new strategy to evaluate the clinical utility of AI-made reports: a straightforward, LLM-based diagnostic evaluation. First, we prompt a zero-shot LLM (Llama 3.1 70B AWQ, prompts in \S\ref{ap:prompts}) to identify in which organ the report mentions tumors. Then, we convert the LLM answer into categorical labels. We compare labels for AI-made and human-made reports (ground-truth) to calculate tumor detection sensitivity and specificity. This evaluation strategy is  \textit{scalable} and practical: with zero-shot inference, it does not need fine-tuning and is easily adaptable to multiple hospitals. Importantly, our strategy produces clinically relevant metrics (detection sensitivity / specificity), which are easy to interpret by clinicians. Here, we limit our evaluation strategy to tumor detection. However, it can be expanded to evaluate other relevant clinical information and diseases beyond tumors---with simple prompt modifications.

\section{Experiment \& Result}\label{sec:result}

\begin{figure}[t]
    \centering
    \includegraphics[width=1\linewidth]{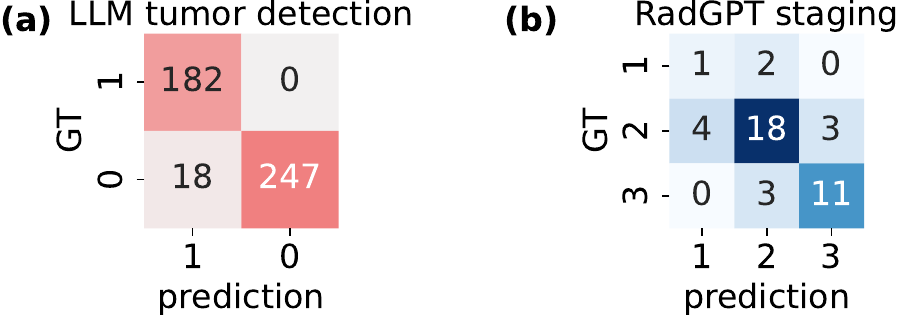}
    \caption{\textbf{Confusion matrices.} \textbf{(a)} A zero-shot LLM (Llama 3.1 70B AWQ) has 96\% accuracy, 0.953 F1-score in determining if radiology reports show tumors. Thus, the LLM can accurately calculate tumor detection sensitivity and specificity for AI-made reports (\S\ref{sec:metric}). LLM's accuracy rivals established labelers, as those in CheXpert \cite{irvin2019chexpert} and CheX-ray14 \cite{wang2017chestx}. Results were manually evaluated by radiologists on 447 reports with kidney, pancreas, and liver tumors. \textbf{(b)} PDAC staging confusion matrix for \method, the first public AI for staging abdominal CT tumors. Results on a private dataset with ground-truth tumor stage annotations ($N$=42).}
    \label{fig:confusion}
\end{figure}


We randomly selected 10\% of \dataset\ as a test set, where we evaluated \textcolor{black}{6 CT report generation models: CT2Rep \cite{hamamci2024ct2rep}, M3D \cite{bai2024m3d}, CT-CHAT \cite{hamamci2024foundation}, Merlin \cite{blankemeier2024merlin}, RadFM \cite{wu2023towards} and RadGPT---as baselines for future work.}
In \dataset, \method\ transforms radiologist-revised tumor masks into reports. In this section, we evaluate \method\ as a fully-automated, segmentation-assisted method, without radiologist revision (\figureautorefname~\ref{fig:rgpt}). We have both AI trained on \dataset\ (CT2Rep and CT-CHAT, \textcolor{black}{see Appendix~\ref{app:ct2Rep} for training details}) and those trained on abdominal CTs in other works (M3D, Merlin, DiffTumor inside \method). To ensure realistic evaluation \cite{bassi2024touchstone,bassi2024improving}, we evaluate on the \dataset\ test set and on a private out-of-distribution (OOD) dataset, from the University of California San Francisco hospital (UCSF, California, USA) never seen by any AI in training.

\textbf{Zero-shot LLMs can accurately evaluate report generation.} For automated evaluation on a large test dataset, we will use an LLM (Llama-3.1) to assess the reports generated by \textcolor{black}{6} AI models (\S\ref{sec:metric}). Before that, radiologists verified the LLM's ability to determine whether a report indicates tumors or not. They \textit{read the zero-shot LLM answers for 447 different reports}, verifying that it achieved 96\% accuracy (\figureautorefname~\ref{fig:confusion}). Results demonstrate the LLM reliability in evaluating tumor detection, per-organ.

\textbf{Segmentation can assist report generation models.} The LLM-based evaluation, (\tableautorefname~\ref{tab:radGPT}) showed that the reports generated by \method\ strongly surpassed the other abdominal CT report generation models, especially in the OOD test set (unseen hospital)\footnote{As \method\ narrative and structured reports match in diagnostic accuracy we present only one result for \method.}. End-to-end trained methods had difficulty detecting tumors in the OOD dataset (low sensitivity), and \method\ strongly outperformed them for small and large tumors in the liver, pancreas, and kidneys. This performance difference shows the benefits of using segmentation to improve report generation: DiffTumor produces accurate tumor segmentations, which \method\ translates into reports. By releasing \dataset, the first abdominal CT dataset with triplets of CT scans, reports, and per-voxel annotations, our objective is to catalyze further research on segmentation-assisted report generation.

\textbf{\method\ is the first public AI model to perform cancer staging on abdominal CT.} \figureautorefname~\ref{fig:confusion} shows the performance of \method\ for staging of pancreatic adenocarcinoma. \method\ fully-automated reports achieved accuracy of 71.43\% in determining tumor T stages 1 to 3. The results show that AI is a promising tool for assisting cancer staging, a key but time-consuming task for radiologists. Still, these fully-automatic results show radiologist revision is essential to ensure staging accuracy in \dataset.

\subsection{RadGPT Accurately Measures Tumor Size}

\begin{table}[t]
\centering
\scriptsize
\begin{tabular}{lccc}
\toprule
 & liver tumor & pancreatic tumor & kidney tumor \\
\midrule
Detection Precision (\%) & 92.3\scalebox{0.5}{ (12/13)} & 50.0\scalebox{0.5}{ (8/16)} & 91.7\scalebox{0.5}{ (11/12)} \\
Size Accuracy (\%)              & 100.0\scalebox{0.5}{ (12/12)} & 75.0\scalebox{0.5}{ (6/8)}  & 100.0\scalebox{0.5}{ (11/11)} \\
\bottomrule
\end{tabular}
\caption{\textbf{\method\ has  75.6\% tumor detection precision and 93.5\% tumor measurement accuracy.} A radiologist manually evaluated reports \method\ created for 23 external test CTs (UCSF). A tumor measurement was considered correct if it deviated by $\leq$10\% from the radiologist’s measurement (both use the WHO measuring standard \cite{miller1981reporting}). As evaluation is time-consuming, the radiologist evaluated 23 reports. Using an LLM for automatically evaluating tumor measurements is challenging: it requires pairing tumors in AI-made reports and ground-truth reports.}
\label{tab:tumor_level_eval}
\end{table}

An expert radiologist manually evaluated structured reports generated by \method. He analyzed each reported tumor, evaluating its measurement and checking if the tumor is a false-positive (tumor not present in the CT volume) or a true-positive (present). The radiologist deemed 75.6\% of the tumors reported by \method\ true-positives, and 93.5\% of them were correctly measured (\tableautorefname~\ref{tab:tumor_level_eval}). \method\ only made measuring mistakes for pancreatic tumors (PDAC), but even the radiologist could not measure 3 PDACs.

\subsection{RadGPT Locates Tumors in Organs}

\begin{figure}[t]
    \centering
    \includegraphics[width=\linewidth]{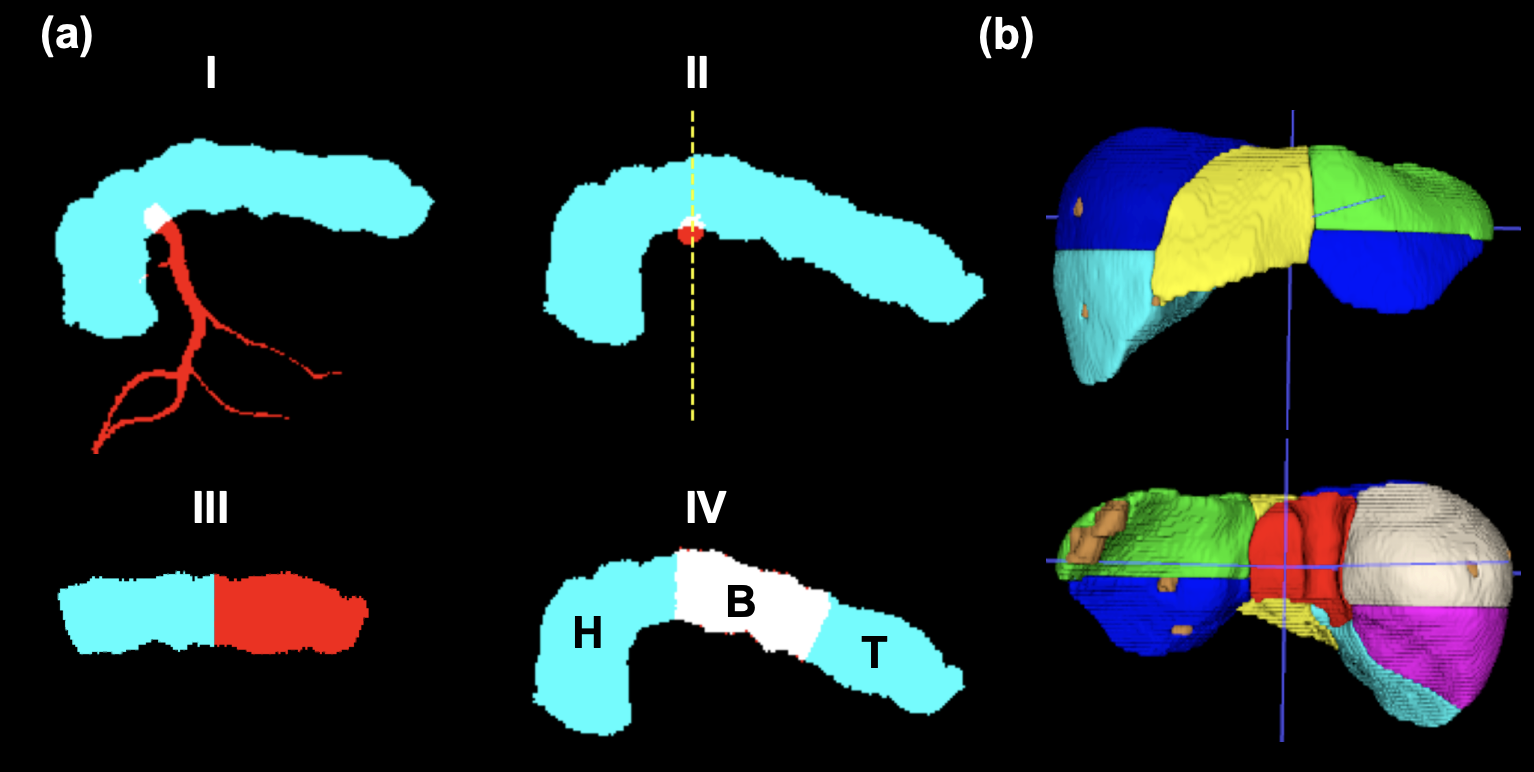}
    \caption{\textbf{Pancreas and liver sub-segments.} \textbf{(a)} \method\ segments the pancreas based on radiology standards \cite{triay2022pancreas}. (I-II) the SMA separates the pancreas head (H) from the body (B), and (III) the remaining pancreas is divided at its midpoint into the body and tail (T). \textbf{(b)} Our liver sub-segmentation model achieved a DSC of 0.85 in segmenting eight liver sub-segments on a public test set \cite{zhang2024robust}. Sub-segments are in different colors and tumors in brown. Sub-segments are essential for \method\ to locate tumors.}
    \label{fig:subsegs}
\end{figure}

\method\ uses use organ sub-segments to locate tumors. It achieved a Dice similarity coefficient (DSC) of 0.85 in segmenting eight liver sub-segments, according to the test set from Zhang~\etal~\cite{zhang2024robust}. For pancreas sub-segmentation, we do not have a ground-truth or dataset for testing, because \dataset\ is the first public dataset to present pancreas sub-segments (head, body, and tail). However, our algorithm to sub-segment the pancreas closely follows radiologist-accepted standards (see \figureautorefname~\ref{fig:subsegs}), and we asked radiologists to qualitatively evaluate our annotations.

\subsection{RadGPT Enhances Human-made Reports}
\label{sec:fusion_results}

Human-made radiology reports often omit critical quantitative details—such as tumor volumes and attenuation (HU) values—compromising clinical decision-making. In our evaluation of 90 human reports from a UCSF, none reported organ or tumor volumes and only 63\% measured all detected tumors. In contrast, our structured reports (\method) consistently provide full, quantitative data. As shown in \tableautorefname~\ref{tab:comparison}, while human-made reports measure volume and HU in 0\% of cases and capture all tumors in only 63\% of cases, \method\ achieves 100\% for all metrics. This level of consistency streamlines clinical assessments and improves prognostic accuracy by ensuring every tumor is precisely measured \cite{woodruff2015tumor, mcerlean2015evaluation, cao2019volumetric}.

In \dataset, 240 CT scans include clinical notes, which lack quantitative tumor measurements; for instance, among 63 TotalSegmentator notes mentioning tumors, none provide such data—even though they report other findings like calcified arterial plaques. By merging these notes with our structured and narrative reports using an LLM, \method\ generates 240 enhanced reports that integrate the notes' comprehensive clinical findings (covering 66 diagnoses) with precise tumor sizes (see \figureautorefname~\ref{fig:fusion_TS}).

\subsection{Discussion: End-to-End, Segmentation-based?}

\tableautorefname~\ref{tab:radGPT} shows \method\ outperformed end-to-end VLMs. The unique design of \method\ offers several \textbf{advantages} over end-to-end training. 
\textbf{\textit{(i)}} Interpretable: \method\ generates reports from tumor segmentation, allowing clinicians and developers to visualize and verify tumor locations and sizes in the CT. In contrast, errors in end-to-end methods are harder to diagnose and debug.
\textbf{\textit{(ii)}} Interactive: Our goal is creating a high-quality dataset of CTs and reports to drive innovation in report generation. As all algorithms make mistakes, a human-in-the-loop approach is key. Segmentation is a \textbf{safeguard}, allowing radiologists to ensure reports are correct, and easily transforming radiologist-revised masks into reports. \textbf{\textit{(iii)}} Strong supervision: tumor segmentation AI has been a long-term focus of the research community, achieving high accuracy by leveraging precise per-voxel masks. Our benchmark shows that segmentation-assisted models can transfer this high accuracy to reports.

\begin{table}[t]
\centering
\scriptsize
\begin{tabular}{lccc}
\toprule
 & Volume & HU & Diameters \\
\midrule
Human-made & 0\%   & 0\%   & 63\% \\
\method\ (ours)    & 100\% & 100\% & 100\% \\
\bottomrule
\end{tabular}
\caption{\textbf{Comparison of human-made reports vs. \method\ reports for 90 UCSF CTs.} Values indicate the percentage of reports containing tumor volume, HU and diameter measurements for all detected tumors. \method\ reports provide more clinically relevant \cite{woodruff2015tumor, mcerlean2015evaluation, cao2019volumetric} quantitative details about tumors.}
\label{tab:comparison}
\end{table}

\section{Conclusion \& Future Work}

Dataset curation and report‑generation are inter‑dependent. 
Developing image-report-mask methods requires image-report-mask datasets, but creating these datasets requires reliable methods and human-in-the-loop involvement. Our focus is creating high-quality image-report-mask datasets to support further methodological advancements.

\dataset\ is the first public dataset providing high-quality abdominal CT scans with reports and per-voxel tumor annotations, encompassing \numofct\ CT scans from \numofhospitals\ institutions. It uniquely includes pancreas sub-segments, peripancreatic blood vessels, and pancreatic cancer stages---absent in existing public datasets.
\method\ transforms per-voxel annotations into structured reports using deterministic algorithms. These reports align with the accuracy of segmentations revised by radiologists in \dataset. Additionally, \method\ enables fully-automated report generation, surpassing existing approaches in detecting tumors.
Together, \dataset\ and \method\ bridge the gap between tumor segmentation and report generation, offering valuable resources and tools to advance AI in abdominal CT interpretation.

We are committed to expanding \dataset\ to include reports for more types of tumors. Additionally, we plan to host benchmarks using \dataset\ with two train/test splits.
\textbf{IID Split:} Randomly sets aside 10\% of the dataset for testing, where training and testing data come from the same institutions, following standard AI evaluation practices. Used in Table \ref{tab:radGPT}.
\textbf{OOD Split:} Uses data from 23 unseen institutions (4,500 CT scans) for testing, providing a large test set to evaluate AI generalization to new environments.
This benchmark will assess report generation models using standard text similarity metrics but will prioritize tumor detection sensitivity and specificity, enabled by our proposed LLM-based diagnostic evaluation.

Although 66 out of 240 fusion reports present diverse diagnoses, \dataset\ is cancer-centric. Cancer is a leading cause of death, and over 40\% of medical imaging reports focus on cancer detection. Thus, AI-assisted report generation has the potential for significant impact. We hope our release of the cancer-centric \dataset\ will stimulate further AI advancements in the field.

\clearpage

\smallskip\noindent\textbf{Acknowledgments.}
This work was supported by the Lustgarten Foundation for Pancreatic Cancer Research and the National Institutes of Health (NIH) under Award Number R01EB037669. We would like to thank the Johns Hopkins Research IT team in \href{https://researchit.jhu.edu/}{IT@JH} for their support and infrastructure resources where some of these analyses were conducted; especially \href{https://researchit.jhu.edu/research-hpc/}{DISCOVERY HPC}. We thank the funding of Italian Institute of Technology and the HPC infrastructure at Italian Institute of Technology. P.R.A.S.B. thanks the funding from the Center for Biomolecular Nanotechnologies, Istituto Italiano di Tecnologia (73010, Arnesano, LE, Italy).

{
    \small
    \bibliographystyle{ieeenat_fullname}
    \bibliography{zzhou,refs}
}

\clearpage
\appendix
\setcounter{page}{1}
\onecolumn
\renewcommand \thepart{}
\renewcommand \partname{}
\part{Appendix} 
\setcounter{secnumdepth}{4}
\setcounter{tocdepth}{4}
\parttoc 

\clearpage
\section{\dataset\ Dataset}

\begin{figure*}[!h]
    \centering
    \includegraphics[width=1\linewidth]{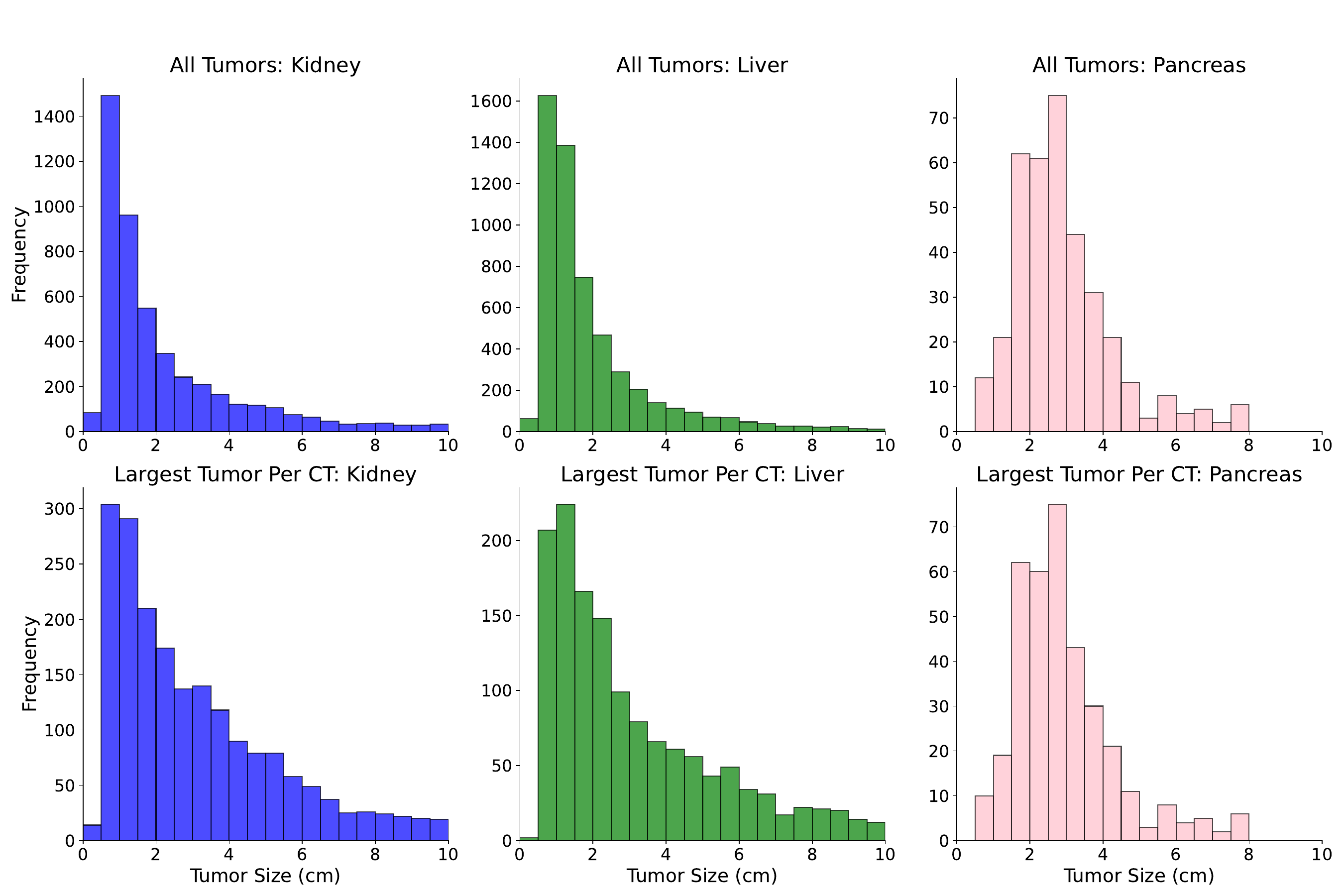}
    \caption{\textbf{Tumor size distribution in \dataset. A large number of CT scans in \dataset\ present small tumors ($\leq$ 2 cm): \textcolor{black}{943}.} The figure’s top row shows histograms of all annotated tumors, while the bottom row focuses on the largest tumor in each organ. Notably, even considering only the largest tumor per organ, \dataset\ still includes a substantial number of small tumors ($\leq$ 2 cm): 504 for kidney, 358 for liver, and 81 for pancreas. These small tumor reports are crucial for training vision-language AI models to detect early-stage cancers, where identifying subtle abnormalities is critical for early detection and treatment.
    }
    \label{fig:tumor_histogram}
\end{figure*}

\clearpage
\subsection{Visualizations}

\subsubsection{Cancer Staging and Blood Vessels}

\begin{figure*}[!h]
    \centering
    \includegraphics[width=1\linewidth]{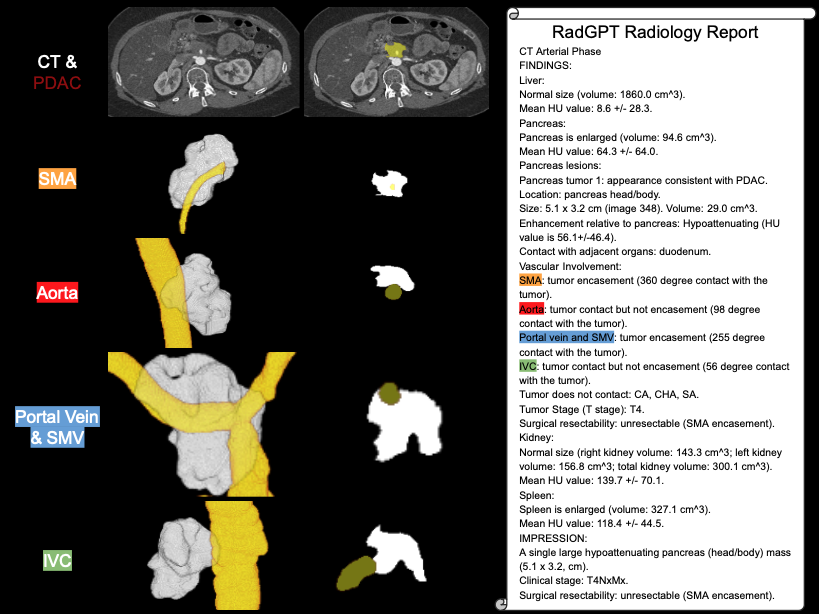}
    \caption{\textbf{Our pancreatic tumor (PDAC) staging report for a stage T4 tumor.} To determine the PDAC T stage, radiologists measure the tumor’s size and evaluate its interactions with critical nearby blood vessels. \method\ automatically replicates this process by utilizing per-voxel annotations of the PDAC and surrounding major blood vessels. The figure highlights these segmentations, and the report shows the angles of contact between the tumor and the blood vessels. In this case, the PDAC fully encases the superior mesenteric artery (SMA), which is a vital vessel supplying blood to the intestines. Surgical removal of a tumor encasing the SMA is not feasible because the artery cannot be reconstructed or bypassed without severe risk to the patient’s survival. This involvement classifies the tumor as surgically unresectable and a stage T4 tumor.
    }
    \label{fig:tumor_histogram}
\end{figure*}

\clearpage
\subsubsection{Pancreas Sub-Segments}

\begin{figure*}[!h]
    \centering
    \includegraphics[width=1\linewidth]{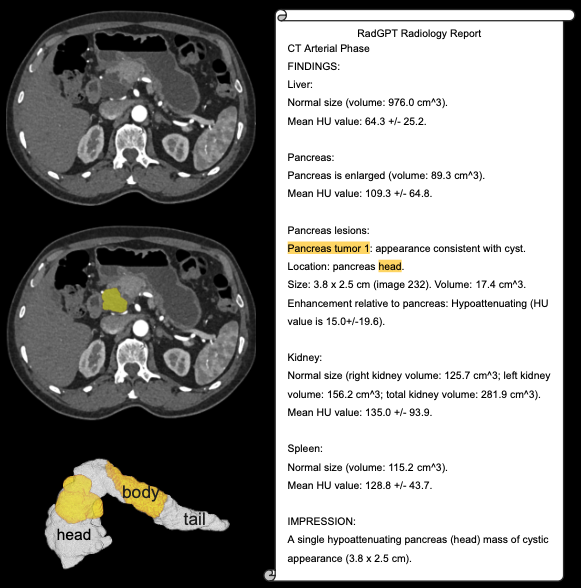}
    \caption{\textbf{CT scan with 2 pancreatic tumors (yellow), and illustration of pancreas sub-segmentation into head (white, left), body (yellow, middle) and tail (white, right).} \method\ used the sub-segments to locate both PDAC tumors in the pancreas head. \dataset\ is the first to present pancreas sub-segments annotated per voxel. This information is crucial for writing radiology reports, as localizing pancreatic tumors in the pancreas head, body or tail is key for determining if the tumor can be surgically removed, and for tracking tumors in time.
    }
    \label{fig:tumor_histogram}
\end{figure*}

\begin{figure*}[!h]
    \centering
    \includegraphics[width=1\linewidth]{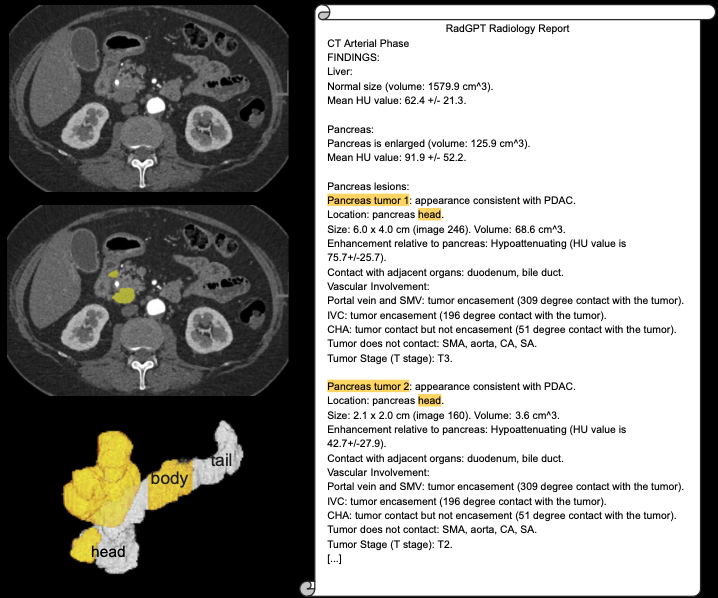}
    \caption{\textbf{CT scan with a pancreatic cyst (yellow), and illustration of pancreas sub-segmentation into head (white, left), body (yellow, middle) and tail (white, right).} \method\ used the sub-segments to locate the cyst in the pancreas head. \dataset\ is the first to present pancreas sub-segments annotated per voxel. This information is crucial for writing radiology reports, as localizing pancreatic tumors in the pancreas head, body or tail is key for determining if the tumor can be surgically removed, and for tracking tumors in time.
    }
    \label{fig:tumor_histogram}
\end{figure*}

\clearpage
\subsubsection{Liver Sub-segments}

\begin{figure*}[!h]
    \centering
    \includegraphics[width=1\linewidth]{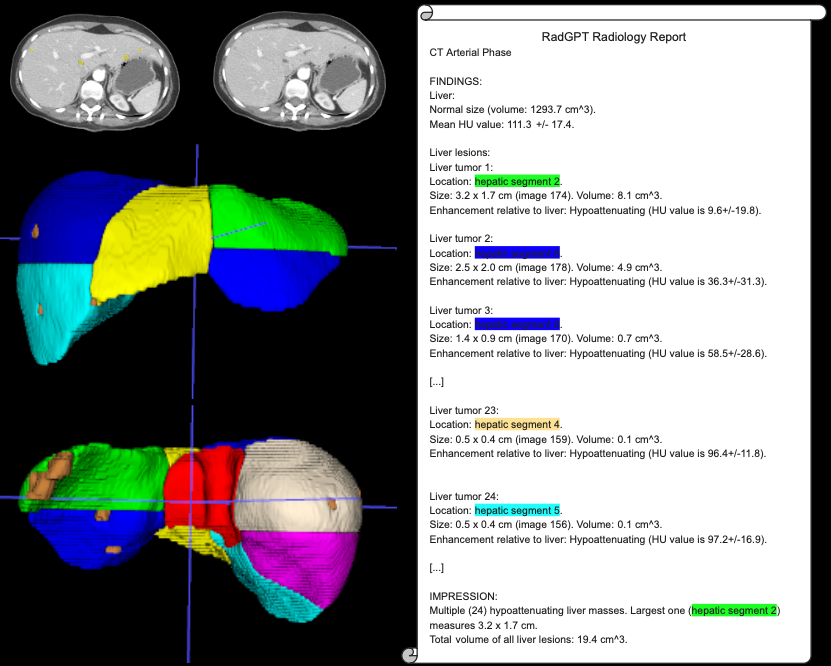}
    \caption{\textbf{CT scan with \textcolor{black}{24} liver tumors (brown), showing how we segment the liver into eight sub-segments for tumor localization.} Notably, unlike our report, most human-made reports would not describe 24 tumors in detail, due to the time required for this task. Liver sub-segments are functionally independent, and can be surgically removed without influencing nearby segments. Thus, localizing tumors into these segments is important for tracking tumors and for surgical planning.
    }
    \label{fig:tumor_histogram}
\end{figure*}

\clearpage
\subsubsection{Kidney Tumor Report}

\begin{figure*}[!h]
    \centering
    \includegraphics[width=1\linewidth]{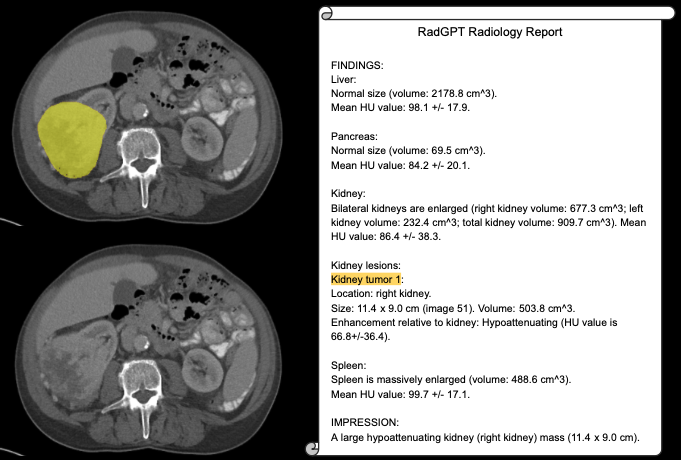}
    \caption{\textbf{CT scan showing a large kidney tumor (yellow) and our report.} 
    }
    \label{fig:tumor_histogram}
\end{figure*}

\clearpage
\subsubsection{Enhanced Human Reports}

\begin{figure*}[!h]
    \centering
    \includegraphics[width=\linewidth]{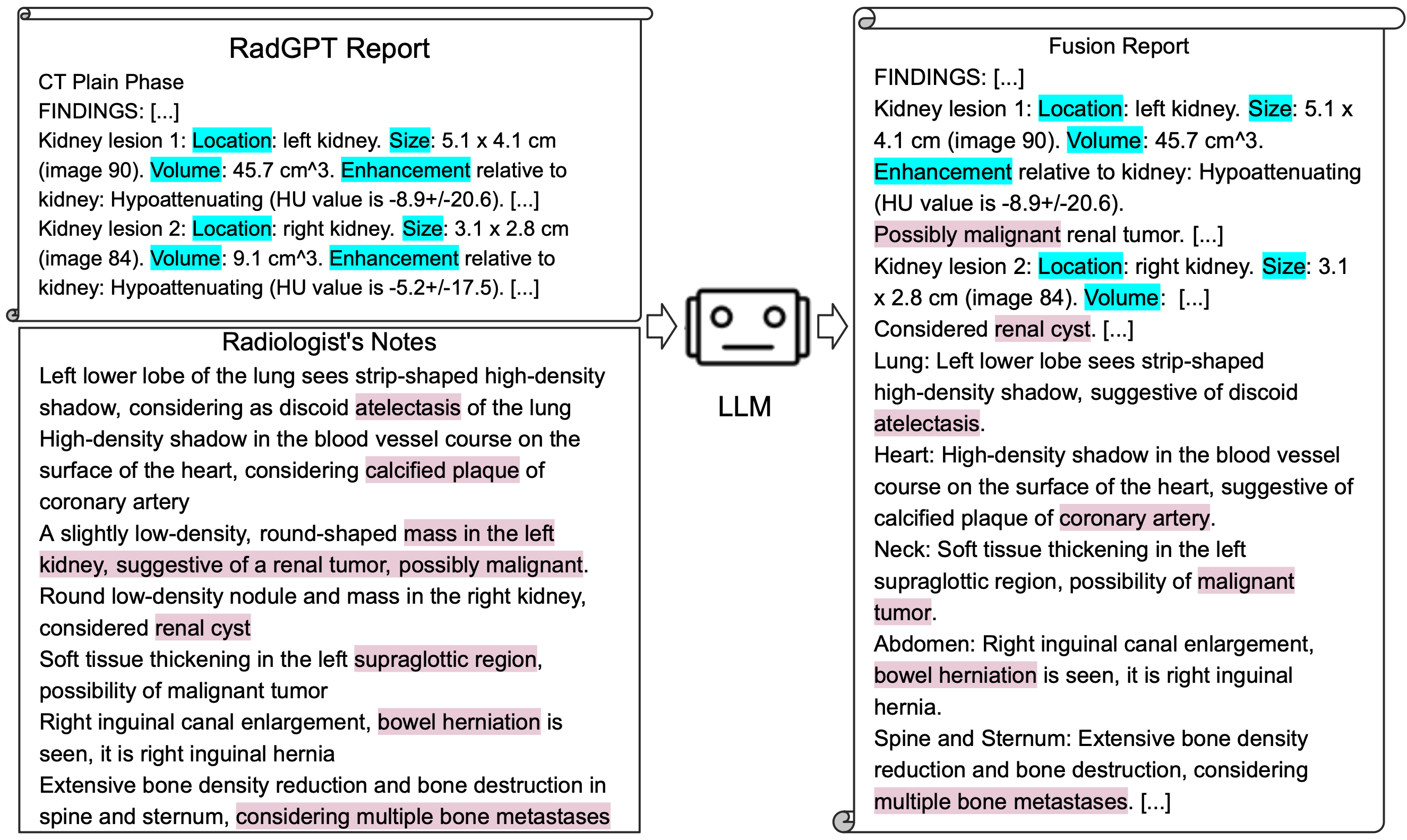}
    \caption{\textbf{In our enhanced human reports, LLMs combine detailed quantitative data from \method’s reports with the generality of human-made reports or clinical notes.} In the image, the LLM extracted kidney tumor sizes and volumes from the \method\ report, while incorporating tumor type and non-cancer-related findings from the radiologist’s notes. Reports created by \method\ include detailed information often absent in human-made reports, such as tumor volumes and individual measurements for a large number of tumors when present. In contrast, human-made reports provide general findings that can expand RadGPT reports to many insights beyond cancer.}
    \label{fig:fusion_TS}
\end{figure*}

\clearpage
\subsubsection{Narrative Reports}

\begin{figure*}[!h]
    \centering
    \includegraphics[width=1\linewidth]{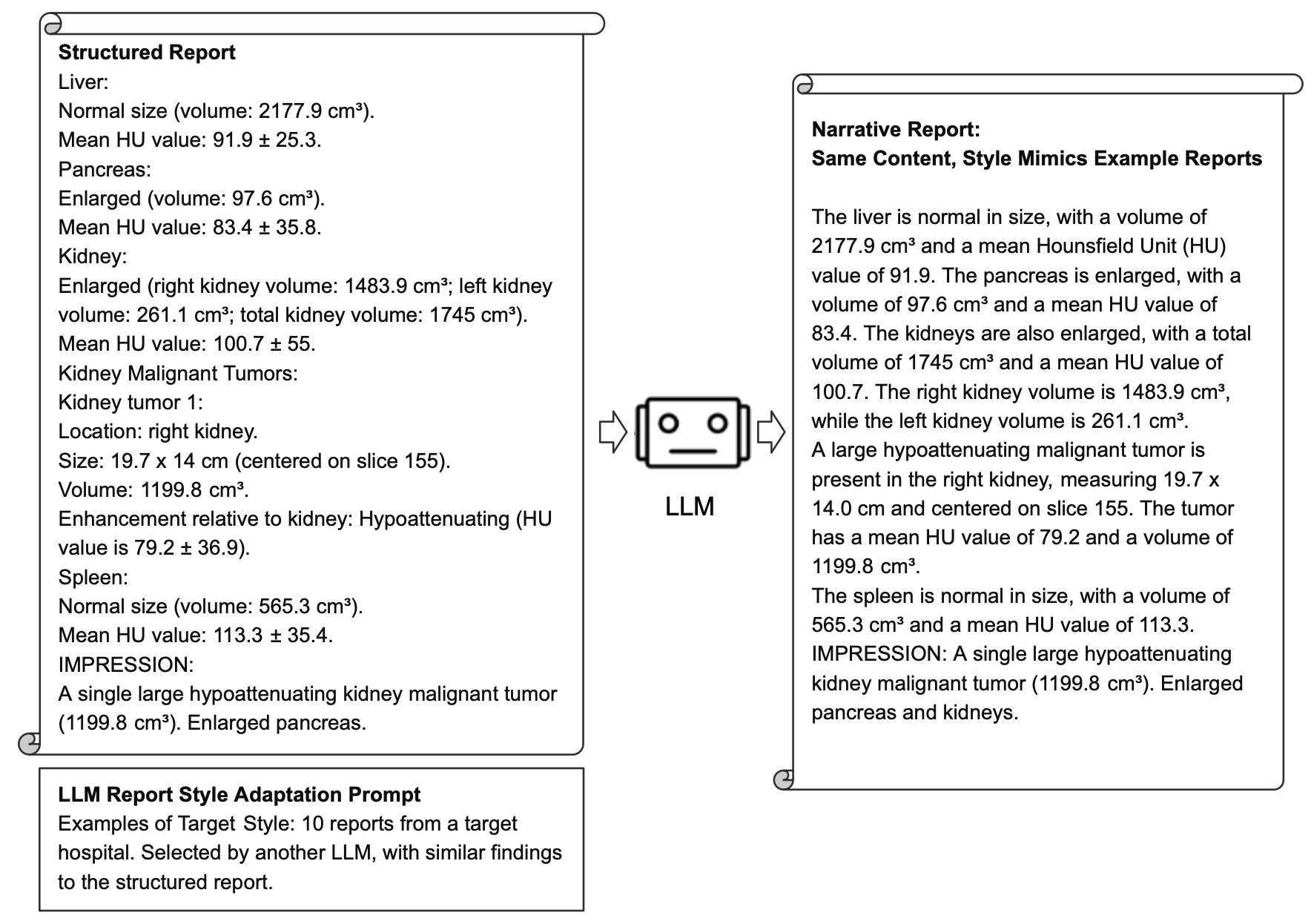}
    \caption{\textbf{Example of Narrative Report: we use LLM to convert structured reports into narrative reports that follow the writing style of a target hospital's.} The LLM receives 10 example reports from the hospitals as examples of style, and is instructed not to change the medical content of the structured report during style adaptation. Since reports targeting diverse abnormalities vary strongly in style, working and structure, we use another LLM to pre-classify the hospital's human-made reports into diagnostic categories (e.g., liver tumor). Thus, during style adaptation, we use as examples only reports that focus on the same diagnosis as the structured report. E.g., if the structured report mentions liver tumors, the examples also will concentrate on liver tumors.
    }
    \label{fig:tumor_histogram}
\end{figure*}

\clearpage
\subsection{Word Cloud}

\begin{figure*}[!h]
    \centering
    \includegraphics[width=0.5\linewidth]{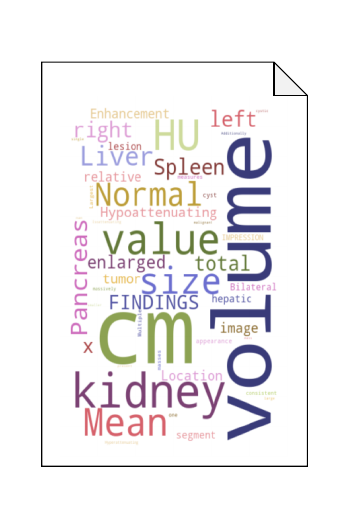}
    \caption{\textbf{Word cloud generated from the \dataset\ reports.} The size of each word reflects its frequency, highlighting the most common terms in the reports. The cloud provides insights about the reports, it clearly shows: their objective nature, focusing on measurements; the inclusion of volumes and HU values, which are usually absent from purely human-made reports; and the focus on cancer and tumor descriptions, with words like tumor, tumor, location, size, and enhancement.
    }
    \label{fig:tumor_histogram}
\end{figure*}

\clearpage
\section{Technical Details of \method}

\begin{algorithm}[h]
\caption{Pancreas Sub-segmentation Using SMA}
\label{algo:pancreas_segmentation}
\begin{algorithmic}[1]
\scriptsize
    \STATE Erase parts of the SMA annotation below the pancreas annotation.
    \STATE Perform PCA on a random subset of the pancreas voxels and rotate the pancreas around its center of mass, aligning its principal component with the x-axis. Rotate the SMA annotation together with the pancreas.
    \STATE Project the SMA onto the x-axis; consider the x-plane at projection's midpoint as the boundary between pancreatic head and body.
    \STATE For the remaining pancreas (excluding head), split body and tail at the x-axis midpoint, using another x-plane.
    
    \FOR{each x-plane (slice) from tail to head}
        \STATE Identify connected components in the current slice.
        \IF{first pancreas slice}
            \STATE Classify all components as body.
        \ELSE
            \STATE Classify components overlapping with the body in the previous slice as body; reclassify others as head. This is important for cases where the pancreatic head bottom crosses the SMA.
        \ENDIF
    \ENDFOR
    
    \STATE Undo rotations and translations; save head, body, and tail segmentations.
\end{algorithmic}
\end{algorithm}

\begin{algorithm}
\caption{WHO-based Tumors Size Measurement}
\label{algo:measurement}
\begin{algorithmic}[1]
\scriptsize
    \STATE Interpolate the tumor segmentation mask to a standard 1x1x1 mm spacing.
    \FOR{each CT slice $s$ containing tumor $A$}
        \STATE Extract the tumor borders by subtracting the tumor segmentation slice $s$ by itself after binary erosion.
        \STATE Calculate the diameter $D_s$ as the longest line between any two points in the tumor borders in $s$.
    \ENDFOR
    \STATE Select the slice $s_{max}$ with the largest diameter $D_{max}$.
    \STATE In the selected slice $s_{max}$:
    \STATE \quad Draw two lines $L_1$ and $L_2$ parallel to the diameter $D_{max}$.
    \STATE \quad Set these two parallel lines $L_1$ and $L_2$ as far as possible from each other while touching the tumor borders.
    \STATE \quad Calculate the distance $d$ between lines $L_1$ and $L_2$.
    \STATE Report the tumor size as $D_{max} \times d$, converting from mm to cm.
\end{algorithmic}
\end{algorithm}

\begin{algorithm}
\caption{Automatic Tumor Staging}
\label{algo:staging}
\begin{algorithmic}[1]
    \scriptsize
    \STATE \textcolor{blue}{\# Make tumor borders overlap with vessels and organs}
    \STATE Apply binary dilation (3x3x3) on tumor mask.
    
    \FOR{each vessel in \{SMA, CHA, CA, SA\}}
        \IF{no overlap with tumor}
            \STATE Set \texttt{contact} = no and continue
        \ENDIF

        \STATE \textcolor{blue}{\# Isolate main vessel branch}
        \FOR{each slice along z-axis from top to bottom}
            \STATE Retain the largest connected component touching the previous slice’s main component, or the largest if within the first 5\% of slices.
        \ENDFOR
        \STATE Apply binary erosion and dilation (5x5x5), overlap with original vessel segmentation, and retain the largest 3D component.

        \STATE \textcolor{blue}{\# Check main branch contact with tumor}
        \IF{no overlap with tumor}
            \STATE Set \texttt{contact} = no and continue
        \ENDIF

        \STATE \textcolor{blue}{\# Align vessel over x-axis and analyze contact with tumor}
        \STATE Skeletonize main branch and align rotate volume, aligning principal component (PCA) with x-axis.
        
        \FOR{each x-coordinate along the x-axis}
            \STATE Check intersection with tumor; if none, continue
            \STATE Align 5mm vessel segment around x-axis using skeleton PCA and crop to 2.5mm
            \STATE \textcolor{blue}{\# Calculate percentage of border contact with tumor to estimate contact angle (vessels are not perfectly round)}
            \STATE Extract vessel borders for each slice and calculate border-tumor overlap percentage
            \STATE Compute contact angle as overlap percentage $\times$ 360; update \texttt{max\_contact} for vessel if new maximum angle is found.
        \ENDFOR
    \ENDFOR

    \STATE \textcolor{blue}{\# Define T stage based on vessel contact and tumor size thresholds}
    \IF{\texttt{max\_contact} for \{SMA, CA, CHA\} $\geq$ 180}
        \STATE Stage = T4
    \ELSE
        \STATE Determine stage by tumor size: T1a $\leq$ 0.5cm, T1b 0.5–1cm, T1c 1–2cm, T2 2–4cm, T3 $>$ 4cm
    \ENDIF
\end{algorithmic}
\end{algorithm}

\subsection{Training CT2Rep \& CT-CHAT \& Merlin on \dataset}
\label{app:ct2Rep}

We trained \textbf{CT2Rep} using only CT scans and structured reports, ignoring the per-voxel annotations in \dataset. Our training strategy for CT2Rep closely followed the code and hyper-parameters published by the model authors \cite{hamamci2024ct2rep}. Possibly, careful search of hyper-parameters and training algorithms for the abdominal region could improve the model's performance. We introduced minimal changes, needed to adapt the model to the abdominal region: we adopted sub-word tokenization to handle decimals frequently present in our reports; we standardized the CT spacing to 1.5 x 1.5 x 1.5 mm, a choice that reduces computational costs while facilitating tumor measurements by maintaining isotropy; to accommodate longer reports, we increased the model's maximum sequence length to 600; and, for hold-out validation (we used 30\% of \dataset\ as the validation set), we used validation loss rather than sequential decoding and BLEU scoring, which significantly reduced validation time. These adjustments, while minimal, were designed to tailor the model for the unique challenges of abdominal CT report generation.

For \textbf{CT-CHAT}, we similarly trained the model using only CT scans and structured reports without using per-voxel annotations from \dataset. While our general training approach again mirrored the original authors' published code and hyper-parameters \cite{hamamci2024foundation}, specific adaptations included standardizing the CT spacing to an isotropic 1.5 x 1.5 x 1.5 mm resolution, selecting a patch size of 20 in each dimension (x, y, and z), and ensuring consistency by center-cropping or padding scans to uniform dimensions of 300 × 300 × 600 mm. Training was performed using four A100 GPUs for 20,000 iterations with a batch size of 16. Furthermore, we employed visual instruction fine-tuning identical to the CT-CHAT authors, using an attention pooling mechanism that reduced tokens generated by CT-CLIP to 256 via learned queries, which were then linearly transformed to match the hidden dimension of the Llama 3.1 8B model. Visual instruction fine-tuning proceeded for 100 epochs.

The training of \textbf{Merlin} also leveraged only CT scans and structured reports, again without incorporating per-voxel annotations from \dataset. While closely aligning with the authors' original published code and hyper-parameters \cite{blankemeier2024merlin}, Merlin required a distinct approach since only the pretrained volume encoder optimized for the abdominal region was available, without the report-generation weights. Consequently, we conducted visual instruction fine-tuning by applying a linear transformation to Merlin’s encoded embeddings, mapping them to the hidden dimension of the Rad LLama2 7B model. This fine-tuning stage continued for 100 epochs, effectively adapting Merlin for abdominal CT report generation.

\subsection {Segmentation Post-processing} 

\begin{figure}
    \centering
    \includegraphics[width=\linewidth]{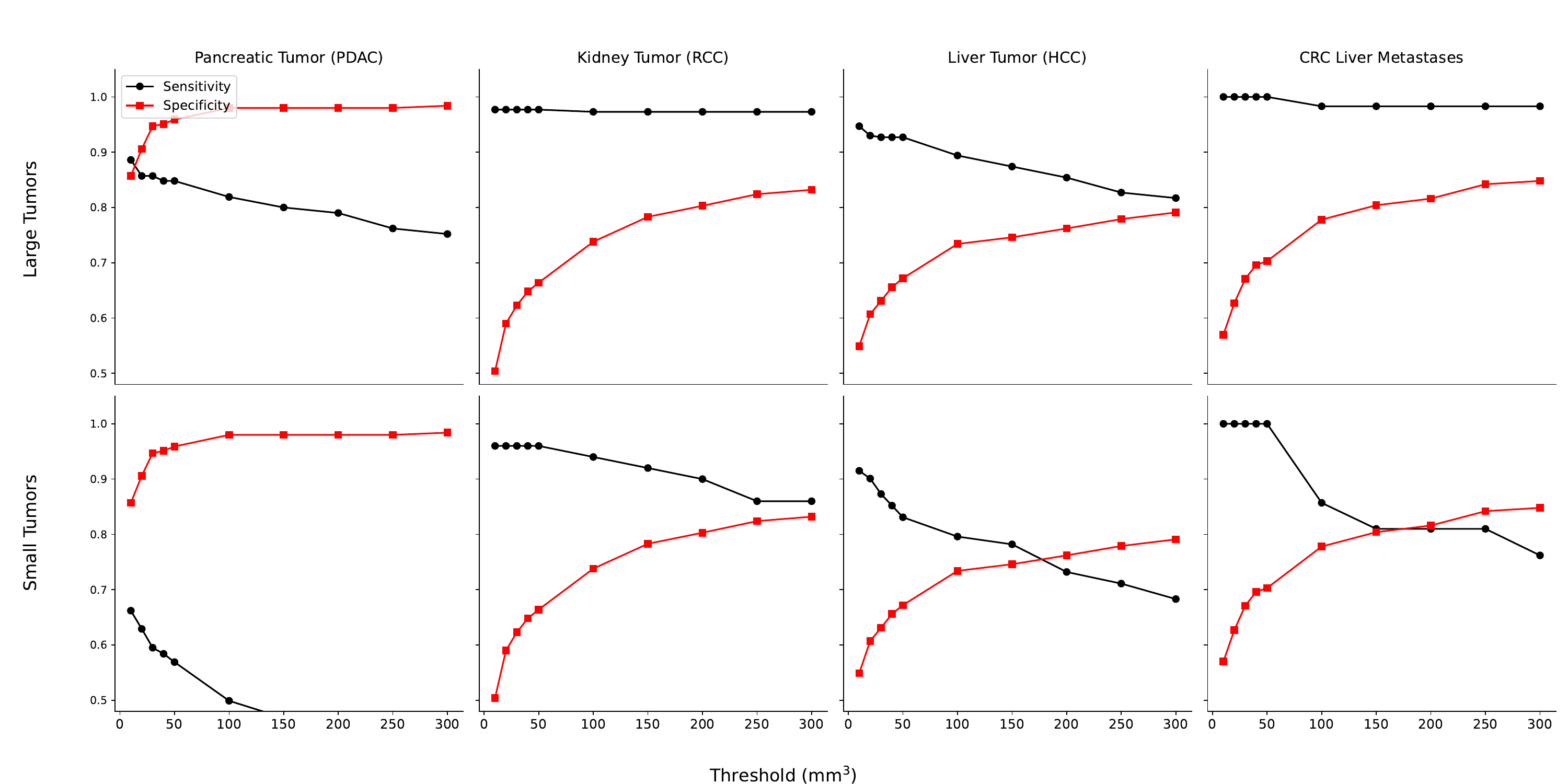}
    \caption{\textbf{Tumor detection sensitivity and specificity for \method\ with diverse thresholds.} Evaluation performed on a private dataset from a hospital never observed during training (UCSF), detailed in \tableautorefname~\ref{tab:radGPT}.}
    \label{fig:th}
\end{figure}

Segmentation models can produce noise: voxels incorrectly labeled as tumors or organs. This may cause false positive cancer detections when \method\ generates reports from nnU-Net or DiffTumor outputs. To address this, we propose a noise reduction algorithm (Alg.~\ref{algo:noise_reduction}). Segmentation noise usually appears as small structures. Thus, we reduce it with binary erosion. Afterwards, to restore the original shape of true tumors and organs, we applied binary dilation followed by a voxel-wise AND with the original tumor segmentation. To further avoid false positives, we perform organ-wise thresholding: we only consider an organ has tumors if the total volume of its tumor voxels is above a small threshold, defined to maximize per-class F1-Score on a validation dataset. For our results section, RadGPT thresholds are: 1 mm\textsuperscript{3} in the pancreas, 150 mm\textsuperscript{3} in the kidneys, 100 mm\textsuperscript{3} in the liver, and 50 mm\textsuperscript{3} for metastases. \figureautorefname~\ref{fig:th} shows specificity and sensitivity for multiple thresholds. Algorithm \ref{algo:noise_reduction} and thresholding are not necessary when we generate \dataset\ reports from radiologist revised segmentations or ground-truth segmentation masks. However, it is recommended when using \method\ without human revision (\figureautorefname~\ref{fig:rgpt}). \figureautorefname~\ref{fig:th} displays performance variation for diverse thresholds.

\begin{algorithm}
\caption{Segmentation Noise Reduction}
\label{algo:noise_reduction}
\begin{algorithmic}[1]
\scriptsize
    \STATE Copy the segmentation output.
    \STATE Apply binary erosion to the segmentation to erase small structures, considered noise. We use a 3x3x3 structuring element, erasing any structure smaller than a 3x3x3 cube.
    \STATE Perform binary dilation on the eroded segmentation. We use a 4x4x4 structuring element.
    \STATE Apply a voxel-wise AND operation between the original mask (before erosion) and the dilated mask, recovering the shape of structures not removed by the binary erosion.
\end{algorithmic}
\end{algorithm}

\subsection{RadGPT Enables Diagnostic Evaluation}
\label{sec:results_metric}

\begin{table}[h!]
\centering
\scriptsize 
\begin{tabular}{@{}lccccccc@{}}
\toprule
 model &  BLEU &  METEOR &  ROUGE-1 &  R.-2 &  R.-L &  BERT &  RadGraph-F1 \\
\midrule
CT2Rep &  0.26 &    8.27 &    12.73 &  0.95 &  7.46 & 45.19 &         3.01 \\
CT-CHAT &  0.15 &    7.19 &    10.90 &  0.54 &  6.00 & 45.63 &         5.69 \\
M3D    &  0.02 &    5.05 &    12.51 &  1.80 &  7.32 & 44.05 &         1.69 \\
Merlin &  0.17 &    7.30 &    11.63 &  0.72 &  6.38 & 45.44 &         \textbf{6.20} \\
RadFM  &  0.10 &    6.90 &    16.14 &  2.41 &  9.26 & \textbf{46.40} &         2.36 \\
\midrule
\method-S & 0.72 &   8.82 &    13.56 & 1.22 & 7.64 & 45.62 &    4.15      \\
\method-N      &  \textbf{0.74} & \textbf{10.47} & \textbf{22.77} & \textbf{2.64} & \textbf{11.89} & 45.41 & 2.92 \\
\bottomrule
\end{tabular}
\caption{\textbf{Report style impacts standard evaluation metrics.} Testing on the same unseen hospital as in \tableautorefname~\ref{tab:radGPT} (UCSF), we evaluate the fully-automated reports from \method\ with text similarity metrics and RadGraph-F1\cite{Yuradgraph} (using RadGraph-XL \cite{delbrouck2024radgraph}, which covers the abdomen region). \method\ narrative (N) and structured (S) reports differ in style only, having the same sensitivity and specificity for tumor detection. However, by mimicking just the style of the test hospital (UCSF, \S\ref{sec:style}), our narrative reports (RadGPT-N) achieved considerably higher  METEOR and ROUGE scores. In contrast, our proposed evaluation metric (\tableautorefname~\ref{tab:radGPT}) only evaluates the diagnosis in the reports and it is not sensible to style variations. 
}
\label{tab:LLM_metrics}
\end{table}

In \tableautorefname~\ref{tab:LLM_metrics}, we use standard text similarity metrics (common in LLM evaluation) and RadGraph-F1 to evaluate the reports generated by the AI. \method\ achieves the highest scores in BLEU, METEOR, and ROUGE. These results align with the superiority of \method\ in our diagnostic evaluation (\tableautorefname~\ref{tab:radGPT}). Thus, diagnostic accuracy may improve LLM metrics. However, BERT and RadGraph-F1 are not aligned with the diagnostic accuracy in \tableautorefname~\ref{tab:radGPT}---RadGPT has considerably superior diagnostic accuracy for cancer (\tableautorefname~\ref{tab:radGPT}), but it does not have the highest BERT and RadGraph-F1 scores. Moreover, NLP metrics are affected by the style of the report. \tableautorefname~\ref{tab:LLM_metrics} includes structured and narrative reports by \method. They contain the same diagnoses and details, but different style (\S\ref{sec:style}), with the narrative reports mimicking the style of the test hospital. In \tableautorefname~\ref{tab:LLM_metrics}, ROUGE \cite{lin2004rouge} was the metric most affected by the style variation. The results show that standard LLM metrics are influenced by style, but the extent of this influence varies between metrics. Conversely, our proposed LLM-based sensitivity and specificity metrics are only influenced by the diagnostic accuracy of the reports. In addition, sensitivity and specificity provide clinicians with an objective and easily interpretable evaluation of AI-made reports, objectively measuring the clinical usefulness of a vision-language model.

\clearpage
\subsection{LLM Prompts}
\label{ap:prompts}
\subsubsection{Style Adaptation}

Or prompt for style adaptation is the following:

"""

You are provided with a \textbf{structured radiology report} and \(n\) other radiology reports that have different writing styles compared to the structured report.

\textbf{Task:} \\
Please \textbf{paraphrase} the structured report to match the writing style of the other reports.

\textbf{Important Guidelines:}
\begin{enumerate}[label=\arabic*., leftmargin=*]
    \item \textbf{Do Not Alter Medical Information:} Do not change, add, or remove any medical details such as tumor measurements, types, or locations. You may remove HU values.
    \item \textbf{Maintain Original Meaning:} Ensure that the rephrased report conveys the same information as the original structured report.
    \item \textbf{Match Writing Style:} Adapt the language, tone, and structure to align with the provided example reports.
    \item \textbf{Begin your report text with \#start and finish it with \#end.}
    \item \textbf{Provide justification:} Go through all medical findings in your rephrased report (e.g., tumor size, no evidence of metastasis) and show where the information comes from in the structured report. Justification should come after \#end.
    \item \textbf{Pay attention to the Example Reports:} Your writing style must be consistent with the examples.
    \item \textbf{Organization must match:} If the examples have an \textit{Impressions} and \textit{Results} section, you must add them. If the example reports talk about all abdominal organs in a single paragraph, you must do so too. You may skip sections you cannot fill due to lack of information, like patient history.
    \item \textbf{Do not add new findings:} If the structured report does not mention the presence or absence of a medical condition (e.g., metastases), you must NOT include it in your rephrased report.
    \item \textbf{Keep coherence:} Avoid going back and forth between medical findings or organs. For example, do not talk about the size of a pancreatic tumor, then mention the liver, and then go back to pancreatic findings. Keep the information about each organ together.
    \item \textbf{Always include an impressions section with the most important findings.}
\end{enumerate}

\textbf{Example of Rephrasing:}

\textit{Structured Report:}
\begin{quote}
PDAC 1: Pancreatic body/tail. Hypoattenuating pancreas PDAC measuring 6.0 x 3.4 cm (centered on slice 356). Its mean HU value is 39.17 +/- 29.65, and its volume is 27.519 cm\(^3\).
\end{quote}

\textit{Paraphrased Report:}
\begin{quote}
\#start

The patient has a pancreatic adenocarcinoma located in the body and tail of the pancreas, measuring 6.0 x 3.4 centimeters (image slice 356). The tumor is hypoattenuating and has a volume of 27.519 cm\(^3\).

\#end

Justification:
\begin{enumerate}[label=\alph*.]
    \item \textbf{Tumor Type:} Maintained as "pancreatic adenocarcinoma", originally "PDAC".
    \item \textbf{Location:} Preserved as "body and tail of the pancreas", originally "Pancreatic body/tail".
    \item \textbf{Measurements:} Kept as "6.0 x 3.4 centimeters", originally "measuring 6.0 x 3.4 cm".
    \item \textbf{Imaging Slice:} Retained as "image slice 356", originally "centered on slice 356".
    \item \textbf{Attenuation:} Maintained as "hypoattenuating", originally "Hypoattenuating pancreas PDAC".
    \item \textbf{Volume:} Kept as "27.519 cm\(^3\)", originally "volume is 27.519 cm\(^3\)".
\end{enumerate}
\textbf{Note:} Removed mean HU value as per guidelines.
\end{quote}

"""

\textbf{Example Reports (Target Style):} \( \{\text{examples}\} \)

\textbf{Structured Report to Paraphrase:} \( \{\text{structured\_report}\} \)

\subsubsection{Enhancing Human Reports}

Our Report Fusion prompt is:

"""

You are provided with a CT scan \textbf{structured radiology report} and notes written by a radiologist, about the same CT scan.

Your task is to identify any information in the notes that is not already included in the structured report and add it to the appropriate sections of the report. Please follow these guidelines:

\begin{enumerate}[label=\arabic*., leftmargin=*]
    \item \textbf{Do not remove} any existing information from the structured report. However, you may improve the report's details using \textbf{only} relevant information from the notes.
    \item \textbf{Avoid adding any new findings} not already mentioned in either the notes or the structured report.
    \item \textbf{Maintain the report's structure.} Carefully place new information in the correct sections inside "FINDINGS", considering which organ the information mentions. For instance, if the notes mention "cirrhosis," add it to the \textbf{"Liver"} section under \textbf{"FINDINGS"}.
    \item \textbf{Add new sections if necessary.} If the notes refer to an organ not covered in the structured report, create a new section for it. If the notes mention patient metadata (e.g., sex and age), you may add it to the beginning of the report.
    \item \textbf{Update the IMPRESSION section if needed.} Besides the FINDINGS, include any critical information from the notes in the report's \textbf{IMPRESSION} section, summarizing or rephrasing it. Do not add new sections if the notes do not provide concrete information for them.
    \item \textbf{Use consistent terminology.} If possible, make the terminology in the sentences you add to the report match the terminology in the original structured report.
    \item \textbf{Begin your report text with \#start and finish it with \#end.}
    \item \textbf{Provide justification.} Explain where in the report you added each piece of information from the notes. Also, explain why other information in the report was not removed or altered.
    \item \textbf{Do not} write non-informative sentences such as "Patient metadata: Not available in the provided notes" or "Sex: Not specified."
\end{enumerate}

The notes are as follows:
\[
\{\text{clinical\_info}\}
\]

The current structured report is:
\[
\{\text{structured\_report}\}
\]
"""

\clearpage
\subsubsection{Labeling/Report Evaluation}

Our prompt is:

\textbf{Instructions:} Discover if the CT scan radiology report below indicates the presence of liver tumors, pancreas tumors, or kidney tumors. Output labels for each of these categories: \textbf{yes} to indicate tumor presence, \textbf{no} for tumor absence, and \textbf{U} for uncertain tumor presence. 

\textbf{Example:} liver tumor presence=yes; kidney tumor presence=U; pancreas tumor presence=no. 

\textbf{Answer with only the labels, do not repeat this prompt.} 

\textbf{Follow these rules for interpreting radiology reports:}
\begin{enumerate}[label=\arabic*., leftmargin=*]
    \item \textbf{'Unremarkable'} means that an organ has no tumor.
    \item Multiple words can describe tumors. Check both the \textbf{findings} and \textbf{impressions} sections of the report (if present) to understand if an organ has tumors. Some words include: metastasis, tumor, tumor, mass, cyst, neoplasm, growth, cancer, index tumor in cancer patients, and tumors listed as oncologic findings.
    \item Consider any tumor, hyperdensity, or hypodensity a tumor, unless the report explicitly states otherwise. Many conditions are not tumors and should not be interpreted as such unless a tumor is also reported along with the disease. Examples include:
    \begin{itemize}
        \item \textbf{Liver conditions:} Hepatitis, Cirrhosis, Fatty Liver Disease (FLD), Liver Fibrosis, Hemochromatosis, Primary Biliary Cholangitis (PBC), Primary Sclerosing Cholangitis (PSC), Wilson's Disease, Liver Abscess, Alpha-1 Antitrypsin Deficiency (A1ATD), Steatosis, Granulomas, Cholestasis, Budd-Chiari Syndrome (BCS), Transplant, Gilbert's Syndrome, ulcers, wounds, infections, inflammations, and scars.
        \item \textbf{Kidney conditions:} Stents, inflammation, postinflammatory calcification, transplant, Chronic Kidney Disease (CKD), Acute Kidney Injury (AKI), Glomerulonephritis, Nephrotic Syndrome, Polycystic Kidney Disease (PKD), Pyelonephritis, Hydronephrosis, Renal Artery Stenosis (RAS), Diabetic Nephropathy, Hypertensive Nephrosclerosis, Interstitial Nephritis, Renal Tubular Acidosis (RTA), Goodpasture Syndrome, and Alport Syndrome.
        \item \textbf{Pancreas conditions:} Pancreatitis, Pancreatic Insufficiency, Cystic Fibrosis (CF), Diabetes Mellitus (DM), Exocrine Pancreatic Insufficiency (EPI), Pancreatectomy, and Pancreatic Pseudocyst.
    \end{itemize}
    \item Examples of specific tumor names include:
    \begin{itemize}
        \item \textbf{Liver:} Hepatic Hemangioma (HH), Focal Nodular Hyperplasia (FNH), Bile Duct Adenoma, Simple Liver Cyst (SLC), Hepatocellular Carcinoma (HCC), Cholangiocarcinoma (CCA), Hepatic Adenoma (HA), Mucinous Cystic Neoplasm (MCN).
        \item \textbf{Pancreas:} Serous Cystadenoma (SCA), Pancreatic Ductal Adenocarcinoma (PDAC), Mucinous Cystadenocarcinoma (MCC), Mucinous Cystadenoma (MCA), Intraductal Papillary Mucinous Neoplasm (IPMN), Solid Pseudopapillary Neoplasm (SPN), Pancreatic Neuroendocrine Tumor (PNET).
        \item \textbf{Kidney:} Renal Oncocytoma (RO), Angiomyolipoma (AML), Simple Renal Cyst, Bosniak IIF Cystic Tumor, Renal Cell Carcinoma (RCC), Transitional Cell Carcinoma (TCC), Wilms Tumor, Cystic Nephroma (CN), Multilocular Cystic Renal Neoplasm of Low Malignant Potential (MCRNLMP), Hydronephrosis, Allograft.
    \end{itemize}
    \item Consider any benign (e.g., cyst) or malignant tumor as a tumor. Thus, any type of cyst is a tumor.
    \item Organs never mentioned in the report have no tumors.
    \item Do not assume a tumor is uncertain unless it is explicitly reported as uncertain. Many words can describe uncertainty, such as: ill-defined, too small to characterize, nonspecific, and uncertain. Reports may express uncertainty about tumor type (e.g., cyst or hemangioma) but still confirm it is a tumor—in this case, consider the tumor a tumor.
    \item Organs with no tumor but other pathologies should be reported as \textbf{no}.
\end{enumerate}

\subsection{Organ size standards}
\label{ap:standards}

Our standards for considering organs as large are based on widely accepted thresholds in radiological and anatomical studies. For the spleen, we consider volumes greater than 314.5 cm³ as large and over 430.8 cm³ as massive, based on thresholds provided by Taylor et al. \cite{taylor1991spleen}. For the kidneys, a volume exceeding 415.2 cm³ for men is considered large, with the threshold adjusted for individual kidneys (half of the total volume) \cite{kilpatrick2018kidney}. Similarly, a liver volume exceeding 3000 cm³ is deemed large, which represents an upper limit for larger individuals, such as a 150 kg man, and highly depends on factors like weight and sex. For the pancreas, volumes above 83 cm³ are classified as large, as per imaging standards discussed by Kondoh et al. \cite{kondoh2018pancreas}.

When size standards depend on variables like weight or sex, we apply thresholds suitable for larger individuals to ensure comprehensive assessments. This approach minimizes the risk of underestimating organ size variations in diverse populations.

\section{Revisions by Radiologists}
\label{ap:radiologist_verification}

In \dataset, organ segmentation masks were created and verified by radiologists through an efficient human-in-the-loop approach \cite{li2024abdomenatlas}. Conversely, the tumor segmentation masks were suggested by AI and radiologists individually verified and corrected them. To verify reports efficiently, radiologists first ensured that structured reports correctly described the already revised per-voxel tumor annotations. This confirmation was key to ensure our deterministic algorithms worked correctly. Then, to verify our narrative and enhanced human reports, we first used our double-check  procedure: an LLM (Llama 3.1 70B) extracted tumor information from the narrative / human enhanced reports and checked if it matched the information in the corresponding structured reports. For cases of mismatch, we prompted the LLM to correct the narrative / human enhanced reports and repeated the double-check. Any remaining mismatch was sent to radiologists. Mismatches were also analyzed to improve our prompts. E.g., radiologists identified that a few narrative or enhanced reports introduced findings absent from the source structured report or human report. To correct this, we started prompting the LLM to justify each finding by quoting sentences in the source structured report or human report. After the double check, we organized structured, narrative and enhanced human reports in a table and radiologists could quickly compare them, confirming they had consistent medical findings.

\section{Detailed Tumor Statistics}

\begin{figure*}[!h]
    \centering
    \includegraphics[width=1\linewidth]{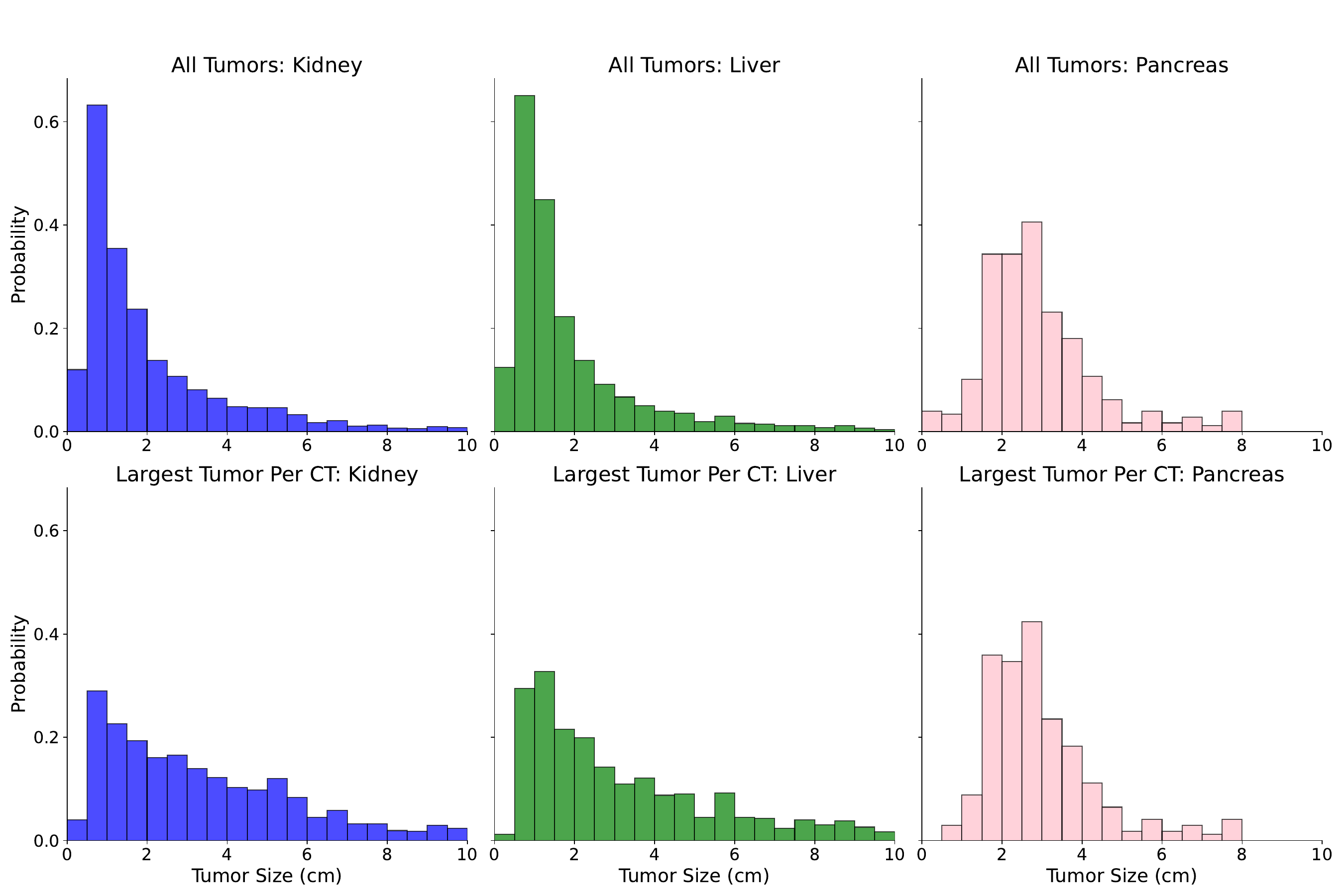}
    \caption{\textbf{Tumor size distribution in \dataset. A large proportion of the CT scans, \textcolor{black}{35\%}, presents small tumors only ($\leq$ 2 cm).} The figure's top row shows histograms of all annotated tumors, while the bottom row focuses on the largest tumor in each organ. Notably, even considering only the largest tumor per organ, the proportion of small tumors ($\leq$ 2 cm) is large in \dataset: 35.59\% for kidney, 38.25\% for liver, and 23.68\% for pancreas. These small tumor reports are vital for training vision-language AI models to detect early-stage cancers, where identifying subtle abnormalities is critical for early cancer detection and treatment.
    }
    \label{fig:tumor_histogram}
\end{figure*}

\begin{figure*}[!h]
    \centering
    \includegraphics[width=1\linewidth]{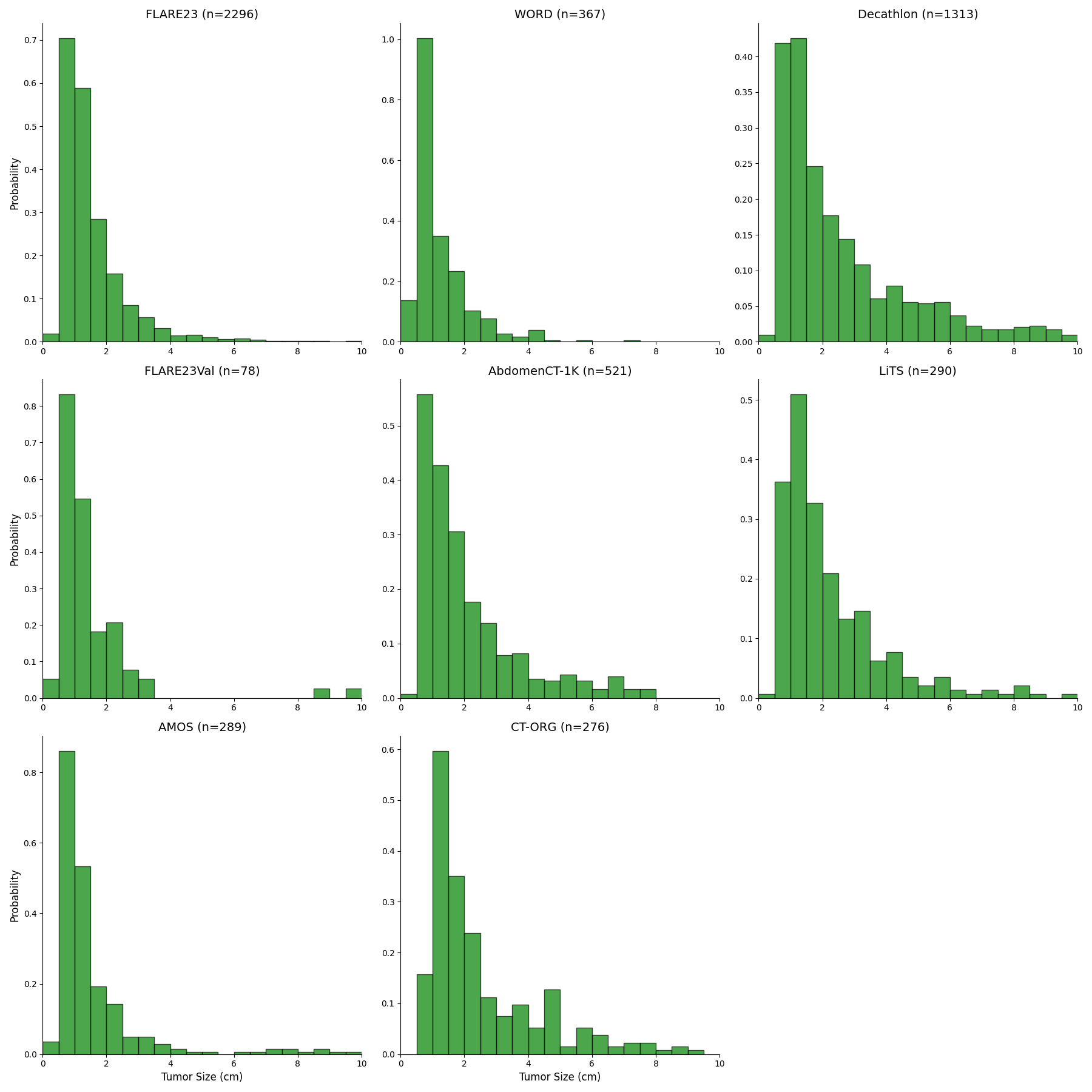}
    \caption{\textbf{Tumor size probability distribution for liver tumors across all datasets in \dataset.} Each subplot represents a dataset with at least three tumor occurrences. The x-axis shows tumor size (cm), and the y-axis represents the probability of tumors within each size range. The figure highlights the variability in tumor sizes annotated across datasets, and the significant presence of small tumors.}
    \label{fig:tumor_histogram}
\end{figure*}
\begin{figure*}[!h]
    \centering
    \includegraphics[width=1\linewidth]{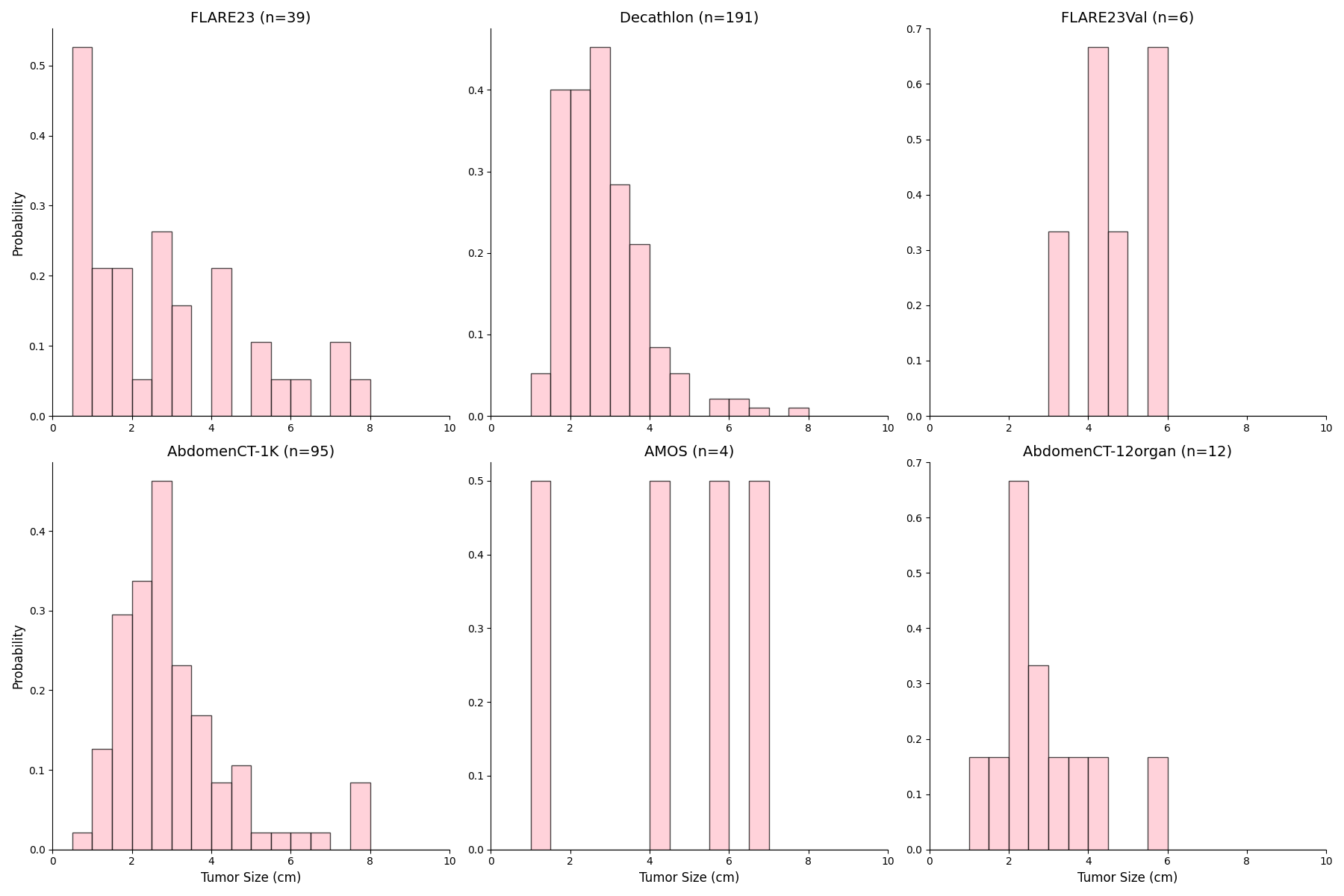}
    \caption{\textbf{Tumor size probability distribution for pancreas tumors across all datasets in \dataset.} Each subplot represents a dataset with at least three tumor occurrences. The x-axis shows tumor size (cm), and the y-axis represents the probability of tumors within each size range. The figure highlights the variability in tumor sizes annotated across datasets, and the significant presence of small tumors.}
    \label{fig:tumor_histogram}
\end{figure*}
\begin{figure*}[!h]
    \centering
    \includegraphics[width=1\linewidth]{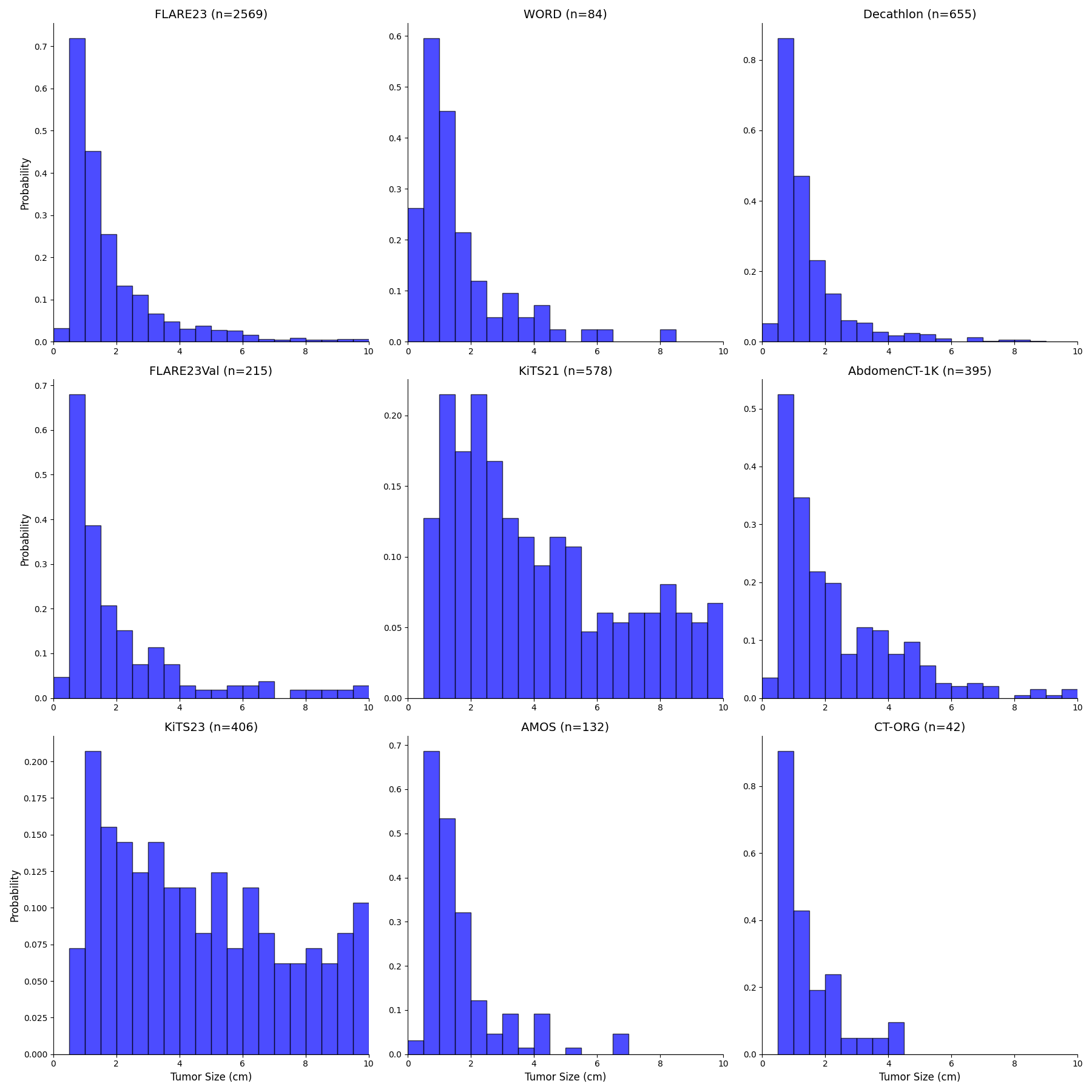}
    \caption{\textbf{Tumor size probability distribution for kidney tumors across all datasets in \dataset.} Each subplot represents a dataset with at least three tumor occurrences. The x-axis shows tumor size (cm), and the y-axis represents the probability of tumors within each size range. The figure highlights the variability in tumor sizes annotated across datasets, and the significant presence of small tumors.}
    \label{fig:tumor_histogram}
\end{figure*}

\begin{figure*}[!h]
    \centering
    \includegraphics[width=1\linewidth]{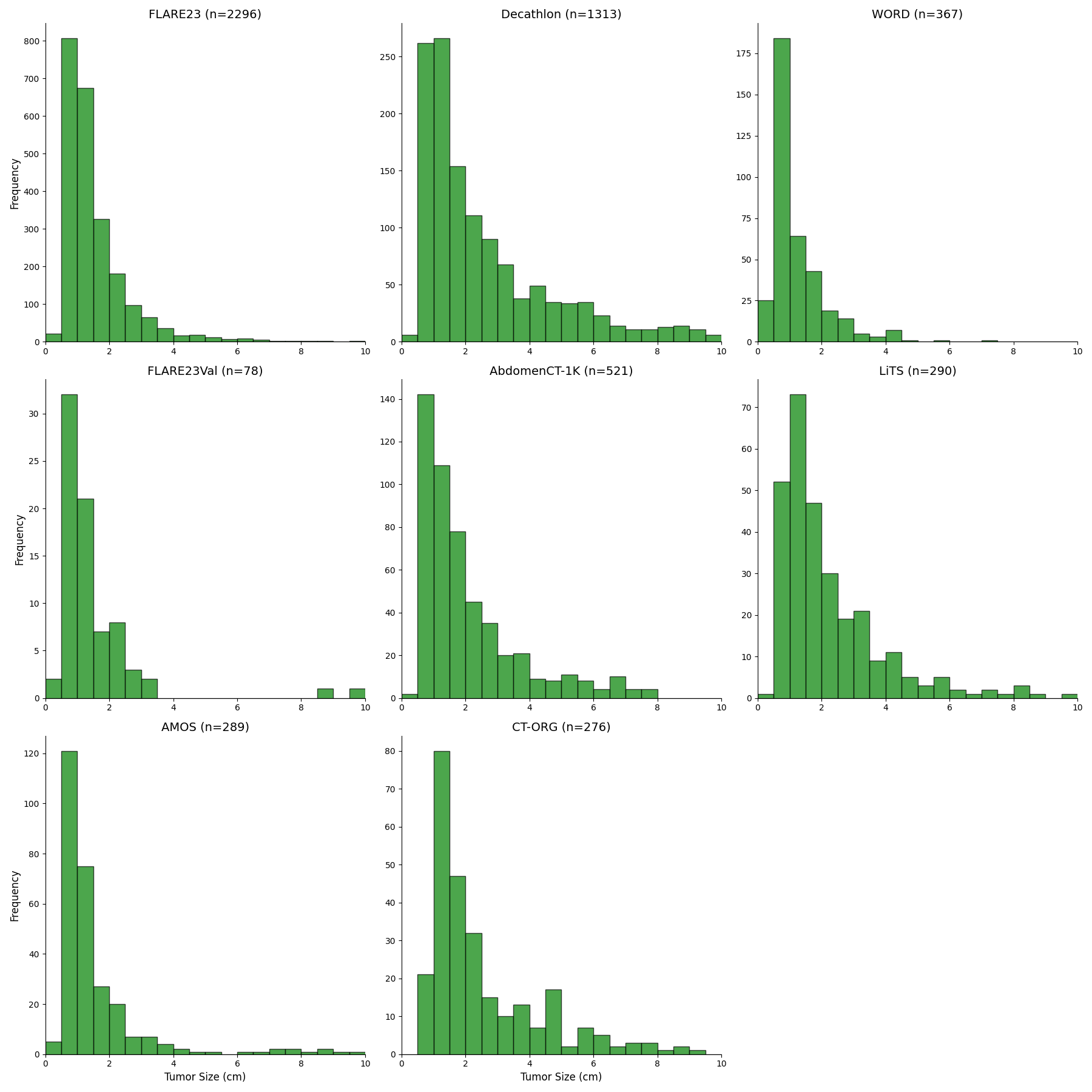}
    \caption{\textbf{Tumor size frequency histogram for liver tumors across all datasets in \dataset.} Each subplot represents a dataset with at least three tumor occurrences. The x-axis shows tumor size (cm), and the y-axis represents the number of tumors within each size range. The figure highlights the variability in tumor sizes annotated across datasets, and the significant presence of small tumors.}
    \label{fig:tumor_histogram}
\end{figure*}
\begin{figure*}[!h]
    \centering
    \includegraphics[width=1\linewidth]{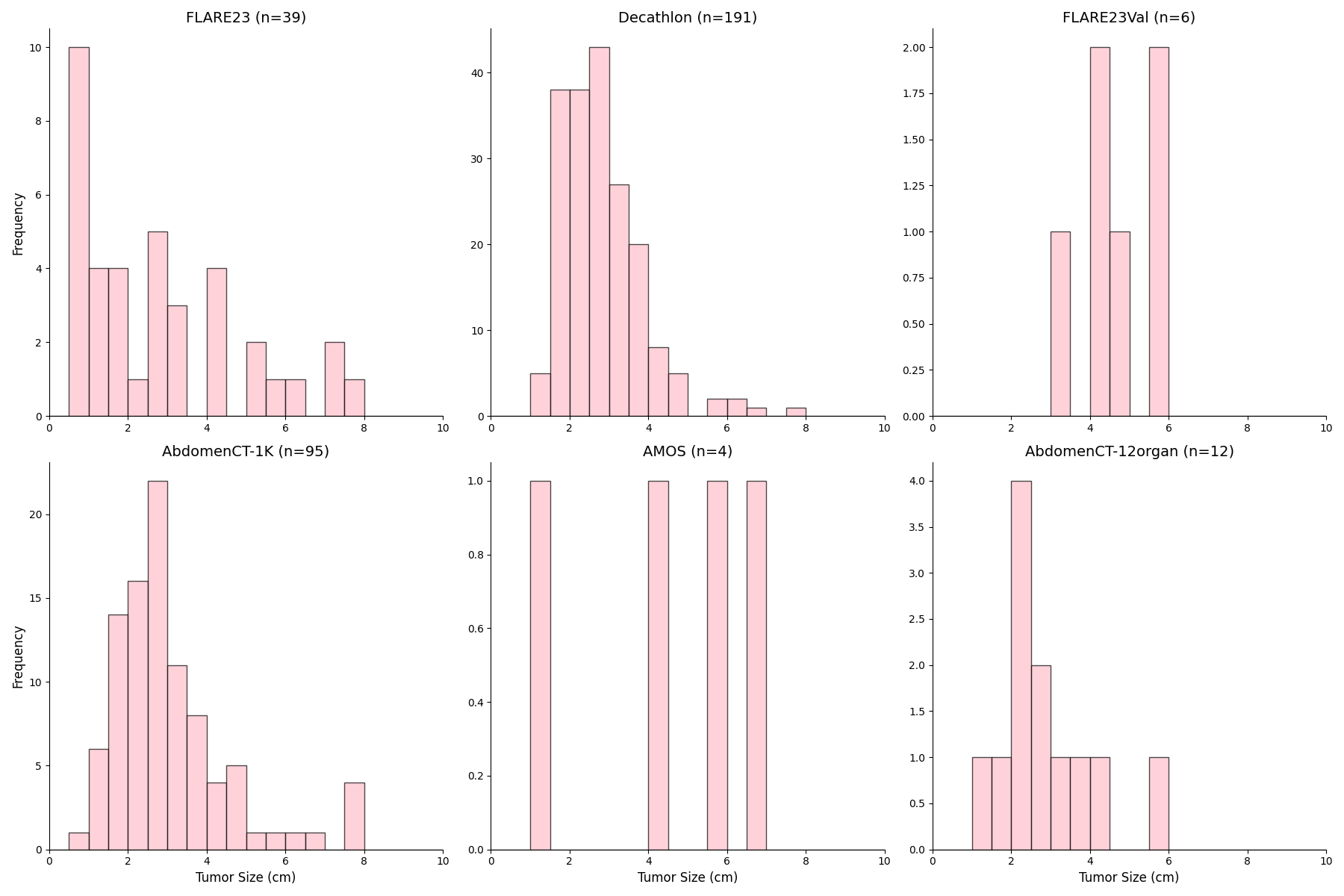}
    \caption{\textbf{Tumor size frequency histogram for pancreas tumors across all datasets in \dataset.} Each subplot represents a dataset with at least three tumor occurrences. The x-axis shows tumor size (cm), and the y-axis represents the number of tumors within each size range. The figure highlights the variability in tumor sizes annotated across datasets, and the significant presence of small tumors.}
    \label{fig:tumor_histogram}
\end{figure*}
\begin{figure*}[!h]
    \centering
    \includegraphics[width=1\linewidth]{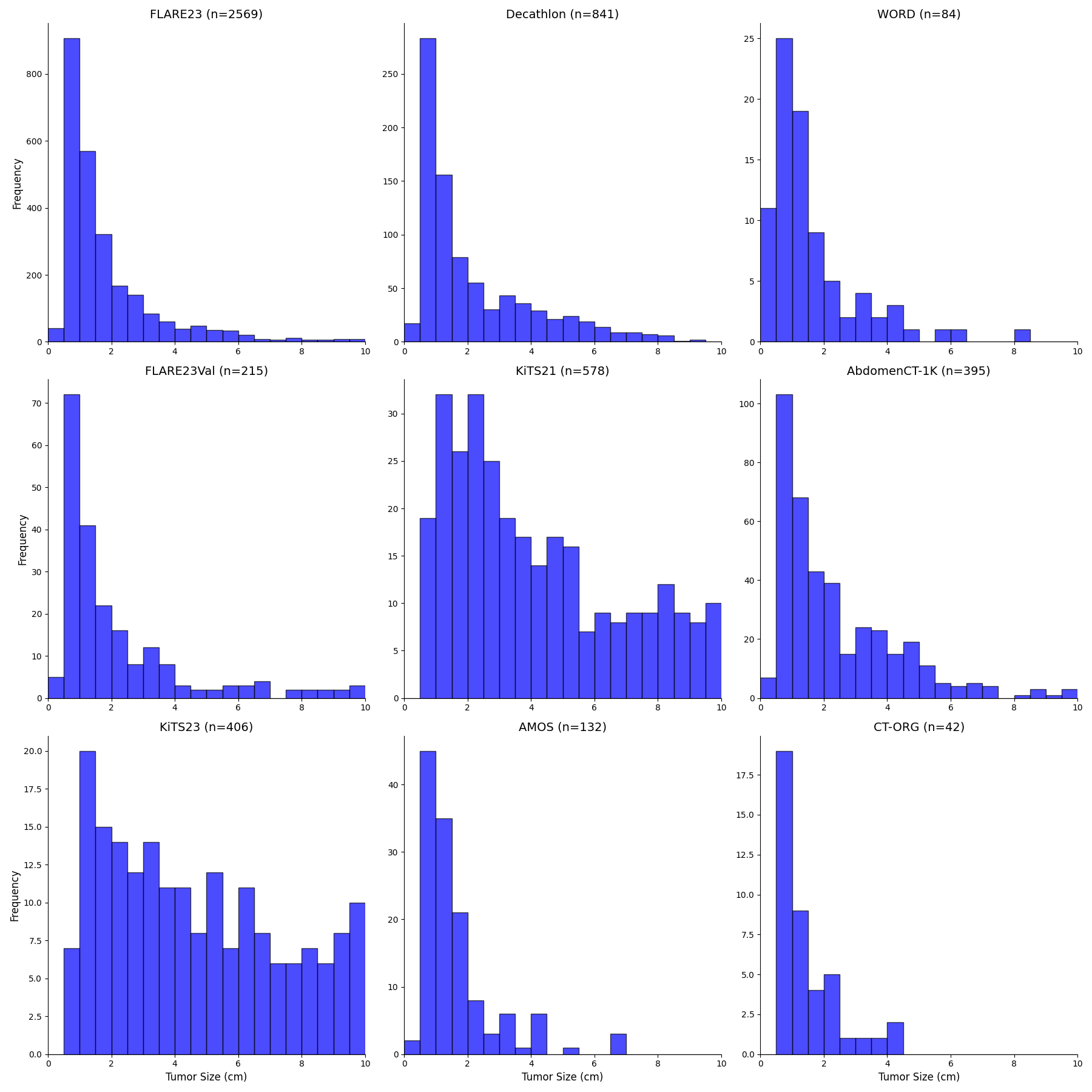}
    \caption{\textbf{Tumor size frequency histogram for kidney tumors across all datasets in \dataset.} Each subplot represents a dataset with at least three tumor occurrences. The x-axis shows tumor size (cm), and the y-axis represents the number of tumors within each size range. The figure highlights the variability in tumor sizes annotated across datasets, and the significant presence of small tumors.}
    \label{fig:tumor_histogram}
\end{figure*}

\end{document}